%% file: main.tex
\documentclass{article}

\usepackage[english]{babel}
\usepackage[letterpaper,top=2cm,bottom=2cm,left=3cm,right=3cm,marginparwidth=1.75cm]{geometry}

\usepackage{amsmath}
\usepackage{graphicx}
\usepackage[colorlinks=true, allcolors=blue]{hyperref}
\usepackage{comment}
\usepackage{amsmath}
\usepackage[makeroom]{cancel}
\usepackage{graphicx}
\usepackage[colorlinks=true, allcolors=blue]{hyperref}
\usepackage{tikzfeynman}
\usepackage{authblk}
\usepackage{cleveref}
\usepackage{todonotes}
\usepackage{amsfonts} 
\usepackage{amssymb}
\usepackage{xcolor}
\usepackage{float}
\usepackage{makecell}
\usepackage{multirow}
\usepackage{tablefootnote}
\usepackage{hyperref}
\usepackage{shuffle}

\date{}
\title{Multiloop calculations with parametric integration\\ in critical dynamics:\\the four-loop analytic study of model A of $\phi^4$ theory}
\author[1,2]{Loran Ts. Adzhemyan}
\affil[1]{Saint Petersburg State University, 7/9 Universitetskaya nab., St. Petersburg, 199034, Russian Federation} 
\affil[2]{BLTP, Joint Institute for Nuclear Research, 
 Dubna, Moscow Region, 141980,  Russian Federation}
\affil[3]{{Petersburg Nuclear Physics Institute named by B.P.Konstantinov of NRC ``Kurchatov Institute'', mkr. Orlova roshcha 1, Gatchina, Leningradskaya oblast, 188300, Russian Federation}}
\author[1,3]{Diana A. Davletbaeva}
\author[1,2]{Daniil A. Evdokimov}
\author[1,2]{Mikhail V. Kompaniets \footnote{m.kompaniets@spbu.ru}}

\begin{document}
\maketitle

\begin{abstract}
We perform an analytical four loop calculation of exponent $z$ in model A of critical dynamics in $d=4-2\varepsilon$ dimensions. This is the first time such a large order of perturbation theory has been calculated analytically for  models of critical dynamics. To do this, we apply the modern method of parametrical integration with hyperlogaritms. We discuss in detail peculiarities of application of this method to critical dynamics, e.g. the problem of linear-irreducible diagrams already present in four loop (contrary to statics where the first linear-irreducible diagram appears in six loop). 
\end{abstract}

Multiloop calculations of Feynman diagrams in quantum field theory (QFT) models are a crucial tool to obtain precise estimates within the perturbative renormalization group (RG) framework. Two main areas studied with QFT methods are high-energy physics and critical phenomena in statistical physics models. The latter encompasses research of various systems such as ferromagnets near the Curie point \cite{ZinnJusti_book,Vasiliev_book,Fisher1998}, vapor-liquid transitions \cite{SHH_vapour76,DL_vapour84,DL_vapour86,AVKK_Hmodel99}, developed turbulence \cite{ForsterNelson_Turb_1977,AVTurbBook1999,AAKVTurb2003,CanetTurb22,AEKTurb2024}, social networks \cite{NW_network99,GGPS_network25,KJBS_network24}, bacterial colony dynamics \cite{MZDGG_bact_20,ZDMGG_bact22}, neural activity \cite{BC_neural07,LW_neural018,TS_neural22,ZHPL_neural25}, collective motion of bird flocks and insect swarms \cite{TT_swarm98,CCGGGP_swarm19,CCGGMOOS_swarm23,T_swarm24}, and more. The principal aim is to identify the existence of critical states and, when they occur, to characterize the associated static and dynamic critical behavior. These behaviors are described by a set of critical exponents - universal numbers governing power-law scaling of physical quantities near criticality. While most exponents pertain to static phenomena (e.g., divergence of the correlation length, correlation function, susceptibility, heat capacity, etc.), several govern purely dynamic effects. The most studied among these is the dynamic exponent $z$, which quantifies the critical slowdown of the relaxation time $\tau$ 
\begin{equation}
    \tau \propto \xi^z\,,
\end{equation}
where $\xi$ is the divergent correlation length of the critical system. 

In this work, we present the four-loop analytic RG calculation of the exponent $z$ in $O(n)$-symmetric model A \cite{Halperin77}, which is a dynamic generalization of the well-known static model $\phi^4$ with purely dissipative relaxation dynamics of the order parameter. Studying this model across different values of spatial dimensions $d$ and order parameter component number $n$  yields varied values of $z$, each corresponding to its own universality class. Initially, model A was applied to describe critical slowdown in ferromagnetic systems with various types of symmetry of the order parameter, such as uniaxial (Ising-like), planar (XY), and isotropic (Heisenberg) cases \cite{AEHIKKZ_5l22,LV_ising23,J_XY00,BHS_XY01,PS_Heisenberg89,AR_heisenberg25}. Since then, multiple lines of evidence have shown that model A applies more broadly, including the continuous ordering transition in specific alloys \cite{L_alloy15}, magnetoelectric dynamics in multiferroic chiral antiferromagnets \cite{NG_multiferroic15}, the superfluid transition \cite{Nalimov_superfluid19,Nalimov_superfluid23}, and even to studies of relativistic heavy-ion collisions \cite{AT_ion25}.

The wide range of physical phenomena described by model A has attracted extensive attention for decades, leading to a plethora of results from experimental studies, numerical simulations, and field-theoretical analyses. A comprehensive review of the modern research on this topic, along with corresponding estimations of the exponent $z$ can be found in \cite{AEHIKKZ_5l_2_22}. Experimental studies are typically associated with significant uncertainties, primarily due to the difficulty of simultaneously maintaining the system at criticality and resolving the dynamics of critical fluctuations. These limitations often result in larger error margins compared to numerical or field-theoretical approaches. Numerical simulations employ a variety of algorithms -- many based on Monte Carlo lattice methods \cite{O_lattice04} - and can achieve high precision, but finite-size effects introduce systematic uncertainties that limit accuracy. Finally, the field-theoretical consideration relies on the RG approach, which allows us to rigorously account for the impact of fluctuations near the critical point. The most established framework is the perturbative RG with the $d=4-2\varepsilon$ expansion, which expresses critical exponents in terms of asymptotic series in the formally small parameter $\varepsilon$. It describes the deviation of the dimension of the space from its upper critical value $d_c=4$, above which the Landau mean field theory is applicable. The calculation of each expansion coefficient requires evaluation of Feynman diagrams with corresponding number of loops. The analytical results were available only up to three loops \cite{AV_3l84}, while the fourth \cite{AIKV_4lSD17} and fifth \cite{AEHIKKZ_5l22,AEHIKKZ_5l_2_22} loops were obtained with the prominent numeric \textit{sector decomposition} (SD) method \cite{BH_sd04} modified for the evaluation of dynamic diagrams \cite{AIKV_4lSD17,AEHIKKZ_5l_2_22,AEHIKKZ_5l22,Perc25}. 

\begin{table}[H]
    \centering
    \begin{tabular}{clcc}
        \hline
        \hline 
        Number of loops & Method & Year & Ref.\\ \hline 
        2 & analytic & 1972 & \cite{HHM_2l2} \\ 
        3 & analytic & 1984 & \cite{AV_3l84} \\ 
        4 & numeric  & 2008 & \cite{ANS_4l08} \\ 
        4 & \makecell{numeric (SD) with  diagram reduction} & 2017 & \cite{AIKV_4lSD17} \\ 
        5 & \makecell{numeric (SD)  with diagram reduction} & 2021 & \cite{AEHIKKZ_5l_2_22} \\  \hline
    \end{tabular}
    \begin{otherlanguage}{english}
    \caption{Summary of the perturbative RG calculations of the dynamic critical exponent $z$ at various loop orders.}
    \end{otherlanguage}
    \label{RG_results}
\end{table}

Due to asymptotic nature of $\varepsilon$ expansion, obtaining a meaningful estimation of the exponent $z$ requires performing a resummation procedure, which is a separate challenging problem. The most straightforward approach is using the Padé approximants method, while a more robust technique is the Borel resummation, which for model A can be refined using known asymptotics of higher orders \cite{HKN_hoa05}. In the five-loop paper \cite{AEHIKKZ_5l22}, the modified \textit{free-boundary condition method} was applied as an advanced novel resummation scheme, as well as \textit{KP17} resummation technique \cite{KP17}, which yields the most precise RG prediction for the exponent $z$: $2.14(2)$ for $d=2$ and $2.0236(8)$ for $d=3$ for model A with $n=1$. We note that our primary goal is to demonstrate the feasibility of applying modern analytic techniques for evaluating Feynman diagrams in dynamic models, so we focus on obtaining an analytical four-loop result (the next to already known analytical three-loop result). Because the five-loop contribution is already calculated numerically, our four-loop expression does not provide additional input to improve existing five-loop resummations; hence, resummation is beyond the scope of this work.

Calculations in models of critical dynamics are known to lag several loop orders behind their static counterparts \cite{Vasiliev_book,IKKN_multi23}. For example, in the static $\phi^4$ model, the current record is the analytical seven-loop calculation of the $\beta$-function and the eight-loop result for the anomalous dimensions \cite{Schnetz_7l22}. This pronounced gap is primarily due to the presence of time dependence in the propagators of dynamic models, which introduces specific complications in diagrammatic evaluations. First, the number of diagrams is significantly larger than in the static case. The standard approach in quantum field theory to manage this complexity is the Integration-By-Parts (IBP) technique \cite{chetyrkin1980}, which reduces the original set of diagrams to a much smaller set of master integrals. However, the IBP method has not yet been fully developed for dynamic diagrams. A significant advance came with the refined four-loop calculation in model A \cite{AIKV_4lSD17}, where the authors introduced a technique called diagram reduction, based  on the equality of some renormalization constants in static and dynamic theories. This method enabled a drastic reduction in the number of dynamic diagrams and was a key step toward the five-loop result, where the number of diagrams was reduced from an initial 1025 to 201 after applying the reduction \cite{AEHIKKZ_5l_2_22}.

Further complication is the presence in momentum representation non-Lorenz invariant propagators $(-i \omega + \lambda k^2)^{-1}$ which result in so-called \textit{time cuts}, 
brought about by integrations over times. 
Moreover, the presence of times as independent parameters implies that even two-point massless diagrams depend not only on external momentum, but also on external frequency. This poses a significant obstacle to the application of analytical calculation methods developed for static models.

Many of the most advanced multiloop results in dynamic models have been achieved with the application of the robust numeric sector decomposition method \cite{ADHIK_multi16,IKKN_multi23,AEHIKKZ_5l_2_22}.  Its ability to evaluate essentially any integral that can be expressed in the Feynman parametrization makes it a reliable brute-force tool. However, as with all numerical approaches, the main drawback lies in optimization: computing a diagram with high precision can be time-consuming. This raises the question whether analytical methods can be adapted for use in dynamic models. One such method - parametric integration with Goncharov polylogarithms (GPLs) or hyperlogarithms \cite{Brown08,Panzer15} \footnote{In \cite{Brown08,Panzer15}, GPLs are referred to as hyperlogarithms, but they denote the same object.} - has proven effective in performing multiloop calculations across a variety of models, including those in particle physics \cite{BPS_hep20, CS_hep22, FL_hep23, LZ_hep22, MP_hep20, AM_hep21} and critical phenomena \cite{KP17,BGKS_phi321,AEKTurb2024}. The key requirement of this approach is a property known as linear reducibility, which ensures that at each step of integration, the integrand can be expressed in terms of GPLs. This condition naturally limits the applicability of parametric integration - for instance, in the $\phi^4$ model, the first non-reducible diagram appears at six-loop order \cite{KP16}. However, recent techniques have been developed that, in some cases, allow linear reducibility to be restored \cite{Panzer_lr14, RatioRoots, HM_LR20}. Given the greater structural complexity of dynamic diagrams, it is reasonable to expect that reducibility issues would arise at lower loop orders than in static models. The first attempt to apply parametric integration with GPLs to a multiloop calculation in a dynamic model was performed in \cite{AEKTurb2024}, where the four-loop analysis of the model of the stochastic fully developed turbulence  in the limit of high spatial dimension was carried out successfully. However, this model was exceptionally simple, and all diagrams turned out to be linearly reducible without requiring any additional manipulations. As a result, the general applicability of the method to dynamic models remained unclear.

The main goal of this work is to test the applicability of parametric integration in the four-loop calculation of model A - one of the simplest models of critical dynamics, yet significantly more complex than the previously studied turbulence model in the $d \to \infty$ limit \cite{AEKTurb2024}. This analysis aims to clarify the prospects of analytical multiloop calculations in the model of critical dynamics in general, and the suitability of the parametric integration approach in particular. 

The paper is organized as follows. Section~\ref{sec:2_Renormalization of the model} describes the renormalization of model A. Section~\ref{sec:3_Diagrammatic technique} reviews the diagrammatic technique and the concept of \textit{time versions} of dynamic diagrams. Section~\ref{sec:4_Diagram reduction} is devoted to the dynamic diagram reduction method, where we present a modification of the original algorithm \cite{AIKV_4lSD17} designed to eliminate divergent subgraphs. In Section~\ref{sec:5_Construction of renormalized integrals} we outline the construction of renormalized integrals, which serves as a prerequisite for the parametric integration method. Section~\ref{sec:6_Parametric integration of dynamic diagrams} focuses on the evaluation of these integrals using \textit{HyperInt}, a \textit{Maple} implementation of the parametric integration algorithm, and introduces a \textit{stream integration} approach for linear irreducible diagrams. Finally, Sections~\ref{sec:7_Calculation of the dynamic exponent z} and \ref{sec:8_Summary} present the calculation of the anomalous dimension and the dynamic exponent $z$, followed by a discussion of the results.

Appendix~\ref{AppendNickel} describes the construction of a Nickel-like notation for enumerating dynamic diagrams. In Appendix~\ref{AppendA} we discuss the origin of additional linear reducibility issues that arise in the case of dynamic diagrams. Appendix~\ref{AppendB} contains explicit expressions for diagrammatic contributions to the renormalization constant. Appendix~\ref{AppendC} provides a derivation demonstrating the equivalence of our three-loop result to the one obtained in \cite{AV_3l84}. Appendix~\ref{AppendD} is devoted to the comparison with the $1/n$-expansion \cite{HHM_2l2}, allowing for partial verification of the analytic results obtained. The Supplementary Material contains the full analytic expressions for all diagrams up to four loops, together with their numerical evaluation up to 200 digits.

\section{Renormalization of the model}  \label{sec:2_Renormalization of the model}
The nonrenormalized action of model A of critical dynamics is defined by a set of two non-renormalized \( n \)-component fields \( \phi_0 \equiv \{\psi_0, \psi'_0\} \) and can be expressed in the form 
\begin{equation}
\label{base_action}
S_0(\phi_0) = \lambda_0 \psi'_0 \psi'_0 + \psi'_0[-\partial_t \psi_0 + \lambda_0 (\partial^2 \psi_0 - m_0^2 \psi_0 - \frac{1}{3!} g_0 \psi_0^3)],
\end{equation}
 where \( \lambda_0 \) is the Onsager coefficient and \( g_0 \) is the coupling constant. Model A is multiplicatively renormalizable. The renormalized action in the spatial dimension \( d = 4 - 2\varepsilon \) reads as follows:
\begin{equation}
\label{renorm_action}
S_R = Z_1 \lambda \psi' \psi' + \psi' [-Z_2 \partial_t \psi + \lambda (Z_3 \partial^2 \psi - Z_4 m^2 \psi - \frac{1}{3!} Z_5 \mu^{2\varepsilon} g \psi^3)],
\end{equation}
where the renormalized fields and parameters are expressed in terms of bare ones as:
\begin{equation}
\label{renorm_params}
\lambda_0 = \lambda Z_\lambda, \quad m_0 = m Z_m, \quad g_0 = g \mu^{2\varepsilon}  Z_g, \quad \psi_0 = \psi Z_\psi, \quad \psi'_0 = \psi' Z_{\psi'}.
\end{equation}
The renormalization constants \( Z_i \) are associated with the renormalization constants from \eqref{renorm_params} by the following relations:
\begin{equation}
Z_1 = Z_\lambda Z_{\psi'}^2, \quad Z_2 = Z_{\psi'}Z_\psi , \quad Z_3 = Z_{\psi'} Z_\lambda Z_\psi,
\end{equation}
\begin{equation}
Z_4 = Z_{\psi'} Z_\lambda Z_m Z_\psi, \quad Z_5 = Z_{\psi'} Z_\lambda Z_g Z_\psi^3.
\end{equation}

Due to the fact that the renormalization constants \( Z_\psi \), \( Z_m \), and \( Z_g \) coincide with their static counterparts, the renormalization constants of the \( \phi^4 \) model \cite{Vasiliev_book}
\begin{equation}
\label{dyn_stat_eq}
Z_\psi = (Z_\psi)_{st}, \quad Z_m = (Z_m)_{st}, \quad Z_g = (Z_g)_{st},
\end{equation}
and because of the existence of the relation \( Z_{\psi'} Z_\lambda = Z_\psi \) \cite{Vasiliev_book}, we can conclude that $Z_1=Z_2$. In fact, we are only interested in \( Z_\lambda \), which is characteristic of this dynamic model:
\begin{equation}
Z_\lambda = Z_1 Z_{\psi'}^{-2} = Z_1^{-1} Z_\psi^2 = Z_2^{-1} Z_\psi^2.
\end{equation}
For our purposes, it is convenient to evaluate it using the renormalization constant \( Z_1 \), which is determined from the diagrams of the one-irreducible Green function \( \Gamma_{\psi' \psi'} = \langle \psi' \psi' \rangle_{1-\text{irr}} / (2 \lambda) \) at zero external frequency \( \omega \) with \( m = 0 \). Henceforth, we use the redefined coupling constant \( u = g S_d / (2\pi)^d \), where \( S_d = 2\pi^{d/2} / \Gamma(d/2) \) is the area of the \( d \)-dimensional sphere of the unit radius. The perturbative expansion of the renormalized Green function reads
\begin{equation}
\label{gamma_expansion}
\Gamma^R_{\psi' \psi'} \big|_{\omega=0, m=0} = Z_1 \left( 1 + u^2 Z_g^2 (\mu / p)^{4\varepsilon} A^{(2)} - u^3 Z_g^3 (\mu / p)^{6\varepsilon} A^{(3)} + \dots \right) = Z_1 \left( 1 + \sum_{i=2}^{\infty} (-1)^{i} u^i Z_g^i (\mu / p)^{2 i\varepsilon} A^{(i)} \right).
\end{equation}
The renormalization constant $Z_1$ in the four-loop approximation in the minimal subtraction (MS) scheme is given by
\begin{equation}
\label{MS}
  Z_1(\varepsilon,u) = 1
    + u^2 \Bigl(\frac{b_{21}}{\varepsilon}\Bigr)
    + u^3 \Bigl(\frac{b_{32}}{\varepsilon^2} + \frac{b_{31}}{\varepsilon}\Bigr)
    + u^4 \Bigl(\frac{b_{43}}{\varepsilon^3} + \frac{b_{42}}{\varepsilon^2} + \frac{b_{41}}{\varepsilon}\Bigr)+ \mathcal{O}(u^5).
\end{equation}
The coefficients $b_{ij}$ are determined by the requirement that all poles in \eqref{gamma_expansion} cancel. They are then expressed in terms of diagrammatic contributions $A^{(i)}$
\begin{equation}
A^{(2)} = \frac{A_{21}}{\varepsilon} + A_{20} + A_{2,-1}\,\varepsilon\ + \mathcal{O}(\varepsilon^2),
\qquad
A^{(3)} = \frac{A_{32}}{\varepsilon^2} + \frac{A_{31}}{\varepsilon} + A_{30}+\mathcal{O}(\varepsilon),
\qquad
A^{(4)} = \frac{A_{43}}{\varepsilon^3} + \frac{A_{42}}{\varepsilon^2} + \frac{A_{41}}{\varepsilon}+\mathcal{O}(1)\,.
\end{equation}
The Green function also involves the renormalization constant for the coupling constant, known from the statics, of which only the first two orders in $u$ are required:
\begin{equation}
     Z_g(\varepsilon,u) = 1 + u (\frac{c_{11}}{\varepsilon})
    + u^2 \Bigl(\frac{c_{22}}{\varepsilon^2} + \frac{c_{21}}{\varepsilon}\Bigr)+ \mathcal{O}(u^3)\,,
\end{equation}
\begin{equation}
     c_{11} =\frac{n+8}{12} \,,\,\,\,\,\,\,\,\, c_{22} = \frac{(n+8)^2}{144} \,,\,\,\,\,\,\,\,\, c_{21} = -\frac{-14+3n}{48}\,.
\end{equation}
In fact, since the coefficients of the highest poles are expressed in terms of coefficients of the first poles, to obtain the exponent $z$ one only needs the coefficient $Z_1^{(1)}$ at the first-order pole in $\varepsilon$ in $Z_1$. The corresponding coefficients $b_{ij}$ are given in terms of $A^{(i)}$ and $c_{ij}$ as
\begin{align*}
b_{21} &= -A_{21}, \\[6pt]
b_{31} &= -A_{31} \;-\; 2\,c_{11}\,A_{20}, \\[6pt]
b_{41} &= -A_{41} \;-\; 3\,c_{11}\,A_{30} \;-\; \bigl(c_{11}^2 + 2\,c_{22}\bigr)\,A_{2,-1}
             \;-\; 2\,c_{21}\,A_{20} \;+\; A_{21}\,A_{20}.
\end{align*}

\section{Diagrammatic technique} \label{sec:3_Diagrammatic technique}
The propagators corresponding to the model \eqref{base_action} can be written in $(k,t)$-representation as follows:
\begin{equation}
\langle \psi(t_1) \psi (t_2) \rangle = \frac{1}{k^2} e^{-\lambda k^2|t_1-t_2|} =\vcenter{\hbox{ 
\begin{tikzpicture}[node distance=1.5cm]
\coordinate (v1); 
\coordinate[right=of v1] (v2);
\arcl{v1}{v2}{0};
\end{tikzpicture}}},
\label{prop_psipsi}
\end{equation}

\begin{equation}
    \hspace{1.1cm} \langle \psi(t_1) \psi' (t_2) \rangle = \theta(t_1-t_2) e^{-\lambda k^2(t_1-t_2)} = \vcenter{\hbox{
    \begin{tikzpicture}[node distance=1.5cm]
    \coordinate (v1); 
    \coordinate[right=of v1] (v2);
    \coordinate(e2) at ($ (v1)!.15!(v2) $);
    \coordinate[above=of e2] (sv1);
    \coordinate[below=of e2] (sv2);
    \arcl{v1}{v2}{0};
    \draw ($ (v2)!.15!0:(sv1) $) -- ($ (v2)!.15!0:(sv2) $);
    \end{tikzpicture}}},
\label{prop_psi'psi}
\end{equation}

\begin{equation}
\hspace{-3.95cm} \langle \psi'(t_1) \psi'(t_2) \rangle = 0 .
\label{prop_psi'psi'}
\end{equation}

The multiplier in front of the diagram is determined by its symmetry and the additional factor $1/2$ in the definition of the one-irreducible Green function $\Gamma_{\psi' \psi'} = \langle \psi' \psi' \rangle_{1-\text{irr}} / 2\lambda.$ The simple exponential dependence on time of the propagators \eqref{prop_psipsi}, \eqref{prop_psi'psi} allows for trivial integration over times. Since the renormalization of the model is considered at $\omega=0$, it is convenient to use the time-versions technique, which makes it possible to represent the result of the integrations over times in the diagrammatic manner \cite{Vasiliev_book}. Being handy for automation, it has been actively used in series of multiloop studies of dynamic models \cite{AVKK_Hmodel99, ADHIK_multi16,AIKV_4lSD17,AEHIKKZ_5l_2_22,AEKTurb2024}. Let us illustrate it with a three-loop example:
\begin{equation}
 \label{3l_tvs}
    \input{tikz_pics/dynamic_som} \bigg|_{\omega=0} = \input{tikz_pics/dynamic_som5} + \input{tikz_pics/dynamic_som3} + \input{tikz_pics/dynamic_som4}
\end{equation}
the numbers $0, 1,2$ represent the times $t_0,t_1,t_2$ corresponding to  each vertex. The presence of the $\theta$-function in the propagator \eqref{prop_psi'psi} results in the sum of time versions, each defined by an ordered sequence of times. Thus, each dynamic diagram can be represented as a sum of its time versions. This is the reason why the number of diagrams in dynamic models significantly exceeds the number of diagrams in static ones.

The result of the integration over times can be schematically shown by dotted lines representing so-called ''time cuts''. We will denote the square of the momentum flowing through a given line as its ''energy'' $E(k)=k^2$. Then, we associate a factor $1/E_{k_i}$ with each solid line $\langle \psi \psi \rangle$, a factor $1$ with each solid line with dash $\langle \psi \psi' \rangle$ and a factor $1/\sum_i E_{k_i}$ with dotted line. The last sum goes over all energies of the diagram lines, which are crossed by the dotted line, where $k_i$ are the momenta of the corresponding lines. For example, the first time version in \eqref{3l_tvs} is expressed as the following
\begin{equation}
\input{tikz_pics/dynamic_som15}
\sim \frac{1}{E_1 E_2 E_4 E_5} \cdot \frac{1}{(E_1 + E_2 + E_3)}  \cdot \frac{1}{(E_1 + E_4 + E_5)},
\end{equation}
other ones can be expressed in the similar manner. The appearance of time-cut propagators, which are characteristic exclusively of dynamic models, significantly complicates the evaluation of diagrams in comparison with the static case.

\section{Diagram reduction for partial elimination of dynamic subgraphs} \label{sec:4_Diagram reduction}

The most challenging aspect of multiloop calculations in dynamic models compared to static models, apart from the presence of the non-standard time-cut propagators, is the significantly larger number of diagrams, corresponding to different time versions. In the study \cite{AIKV_4lSD17}, the authors proposed a method called ''diagram reduction'' which allows to drastically decrease the number of the diagrams. It is based on the equality of some renormalization constants in statics and dynamics \eqref{dyn_stat_eq}, which results in the formulation of reduction rules:
\begin{equation}
    \label{reduction_rules}
    \vcenter{\hbox{\input{tikz_pics/rule4}}} \quad + \quad \vcenter{\hbox{\input{tikz_pics/rule3}}} \quad + \quad \ldots \quad + \quad \vcenter{\hbox{\input{tikz_pics/rule2}}} \quad + \quad \vcenter{\hbox{\input{tikz_pics/rule1}}} \quad = \quad \vcenter{\hbox{\input{tikz_pics/rule5}}}
\end{equation}
where circles denote arbitrary subgraph, which is the same for each graph, and there may be $k_1 \geq 2$ lines to the left of the subgraph and $k_2 \geq 0$ lines to the right. For the dynamic diagrams contributing to $\Gamma_{\psi \psi'}$, these rules allow a complete expression of them in terms of a combination of static ones. For the considered function $\Gamma^R_{\psi' \psi'}$ \eqref{gamma_expansion}, the diagrams can be reduced only partially, eliminating some of the dynamic vertices. Nevertheless, this technique still substantially simplifies the calculations. For instance, in the four-loop study of model A \cite{AIKV_4lSD17}, the authors were able to decrease the overall number of diagrams from 66 to 17; and in the five-loop study \cite{AEHIKKZ_5l_2_22}, from 1025 to 201.

In the aforementioned studies, the diagrams were evaluated with the \textit{sector decomposition} method \cite{BH_sd04} - a powerful and reliable numeric technique which allows to explicitly calculate the coefficients of the poles in $\varepsilon$ of divergent integrals. Within this approach, there is no need to regularize the integrals and perform any subtractions of divergent subgraphs, so their presence did not pose any additional problems. As a result, the main useful outcome of applying the reduction was the decrease in the number of diagrams. In this paper, we use the parametric integration with GPLs as a main calculation tool, which can only be applied for convergent integrals, requiring the subtraction of subdivergencies, both static and dynamic (see the next Section~\ref{sec:5_Construction of renormalized integrals}). {For this reason, we employ a modified version of the reduction algorithm, designed to replace dynamic subgraphs with static ones whenever possible, thereby reducing the number of necessary subtractions.} This is because static subgraphs, which have only one external momentum (all 2-point subgraphs and some 4-point subgraphs), reduce in the massless theory to a known power of that momentum multiplied by a factor whose pole structure is explicit; that factor can be extracted from the full diagram as a common multiplicative term. Thus, the subgraph is replaced by the known momentum power and does not produce a divergence.

Due to the relatively small number of diagrams in the four-loop approximation, the new reduction is performed manually by analyzing the structure of the diagrams. However, to advance to higher orders, it will be necessary to formulate a rigorous algorithm suitable for automation.

\vspace{0.3cm}
To illustrate the modified reduction technique, we consider its application for three- and four-loop diagrams. The set of dynamic subgraphs appearing in these diagrams, which are subject to elimination, is presented in the figure below.
\begin{figure}[H]
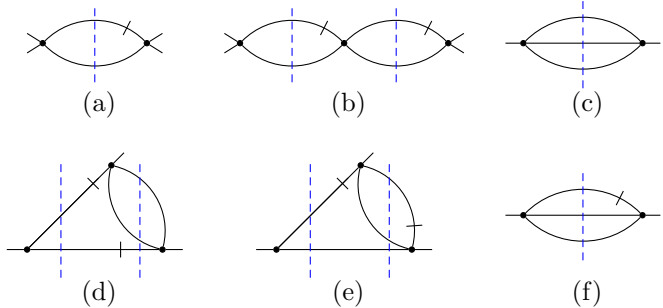

    \centering
    
        \begin{tabular}{c c c}
           \input{tikz_pics/dynamic_loop4}  & \input{tikz_pics/double_dynamic_loop} &
           \input{tikz_pics/dynamic_arbuz8}
           \\
            (a) & (b) & (c) \\ \\
            \input{tikz_pics/glaz1} & \input{tikz_pics/glaz2} & \input{tikz_pics/dynamic_arbuz9} \\
            (d) & (e) & (f)
        \end{tabular}
    
    \caption{The six types of dynamic subgraphs appearing in the the three- and four-loop diagrams. (a)-(e) are logarithmically divergent, (f) is quadratically divergent.}
    \label{subgraph_types}
\end{figure}
\noindent {Let us note that the new reduction procedure may lead to the appearance of static subgraphs that were not present after the original reduction.} However, due to the simpler structure of the static subtractions, it is advantageous to intentionally replace a dynamic subgraph with a static one. 

\vspace{0.3cm}

In the three-loop case, we consider a set of two reduced diagrams obtained from \eqref{3l_tvs} in \cite{AIKV_4lSD17}:
\begin{figure}[H]
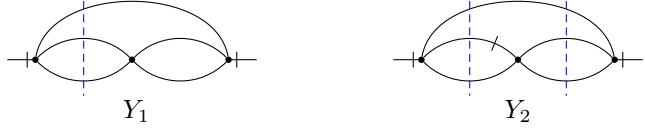

    \centering
    \setlength{\tabcolsep}{24pt}
    \begin{tabular}{c c}
      \input{tikz_pics/Y1}   &  \input{tikz_pics/Y2}\\
       $Y_1$  & $Y_2$
    \end{tabular}
    
    \caption{The set of three-loop reduced diagrams of $\Gamma^R_{ \psi' \psi'}$ from \cite{AIKV_4lSD17}.}
    \label{3l_reduced_orig}
\end{figure}

\noindent The first diagram is free of dynamic subdivergencies, and the second one has a subgraph of type (a) from Fig.~\ref{subgraph_types}. To remove it, we introduce a new diagram 
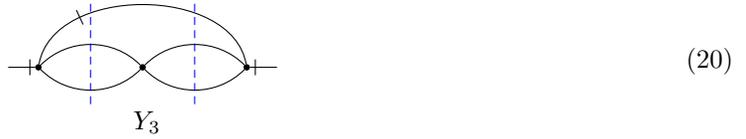
\begin{equation}
\label{3l_new_reduced}
    \begin{tabular}{c}
        \input{tikz_pics/Y3} \\
        $Y_3$ 
    \end{tabular}
\end{equation}
which has no subdivergencies. Using the second rule in \eqref{reduction_rules} and taking into account the symmetry of the diagrams, we obtain the following relation:
\begin{equation}
    Y_1 = 2\,\,Y_2 + Y_3\,\,.
\end{equation}
Consequently, instead of computing the set ($Y_1$, $Y_2$), one can consider the set ($Y_1$, $Y_3$), which contains no dynamic subgraphs.

\vspace{0.3cm}

For the reduction of four-loop diagrams, we begin by considering the 17 reduced diagrams from \cite{AIKV_4lSD17}, listed in Figure 2. They are grouped by four topologies, labeled A, B, C and D. The main goal is to eliminate as many divergent dynamic subgraphs as possible. For demonstration purposes, we restrict our detailed discussion to the reduction of topology A, which consists of the following three diagrams:
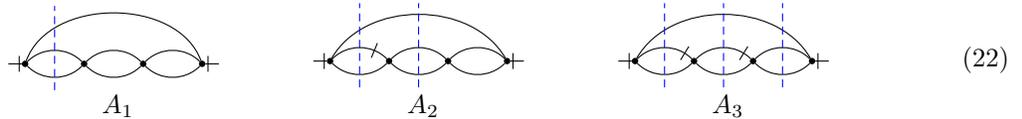
\begin{equation}
    \setlength{\tabcolsep}{16pt}
    \begin{tabular}{c c c}
        \input{tikz_pics/A1} & \input{tikz_pics/A2} & \input{tikz_pics/A3} \\
        $A_1$ & $A_2$ & $A_3$
    \end{tabular}
\end{equation}

The diagram $A_2$ contains one one-loop dynamic subgraph; diagram $A_3$ contains two one-loop and one two-loop dynamic subgraphs; and $A_1$ is free of dynamic subdivergences. We introduce in advance three new diagrams that will be required to perform the reduction:
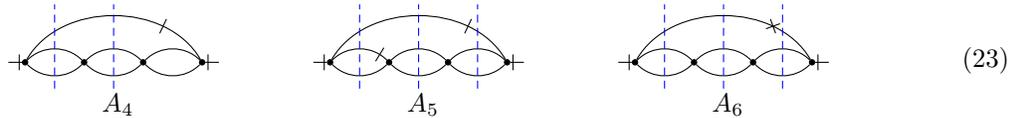
\begin{equation}
    \setlength{\tabcolsep}{16pt}
    \begin{tabular}{c c c}
        \input{tikz_pics/A4} & \input{tikz_pics/A5} & \input{tikz_pics/A6} \\
        $A_4$ & $A_5$ & $A_6$
    \end{tabular}
\end{equation}

To eliminate the subgraph in $A_2$, we consider the diagram $A_4$, which contains no dynamic subdivergences. Due to the symmetry of $A_2$, the following relation holds:
\begin{equation}
    A_1 = 2A_2+A_4.
\end{equation}
Thus, instead of calculating $A_2$, which requires subtraction of the dynamic subgraph, we can calculate $A_4$, which does not require any dynamic subtraction. Similarly, to reexpress $A_3$, we use the diagram $A_5$ with only one subgraph instead of initial three. The corresponding diagram relation is
\begin{equation}
    A_2 = 2A_3+A_5\,\,.
\end{equation}
To remove the last remaining subgraph in $A_5$, we present a modified version of the second relation in \eqref{reduction_rules}:
\begin{equation}
    \vcenter{\hbox{\input{tikz_pics/rule9}}} \quad + \quad \vcenter{\hbox{\input{tikz_pics/rule8}}} \quad + \quad \ldots \quad + \quad \vcenter{\hbox{\input{tikz_pics/rule7}}} \quad + \quad \vcenter{\hbox{\input{tikz_pics/rule6}}} \quad = \quad \vcenter{\hbox{\input{tikz_pics/rule10}}}
\end{equation}
where the cross mark denotes an inverse propagator (without affecting the structure of the time-cut factor). Using this, we introduce the diagram $A_6$ and write the relation
\begin{equation}
    A_4 = 2A_5 + A_6\,\,.
\end{equation}
As a result, after performing the additional reduction, we have managed to obtain the new set of diagrams $A_1$, $A_4$, $A_6$, which does not contain any dynamic subdivergencies.
\begin{figure}
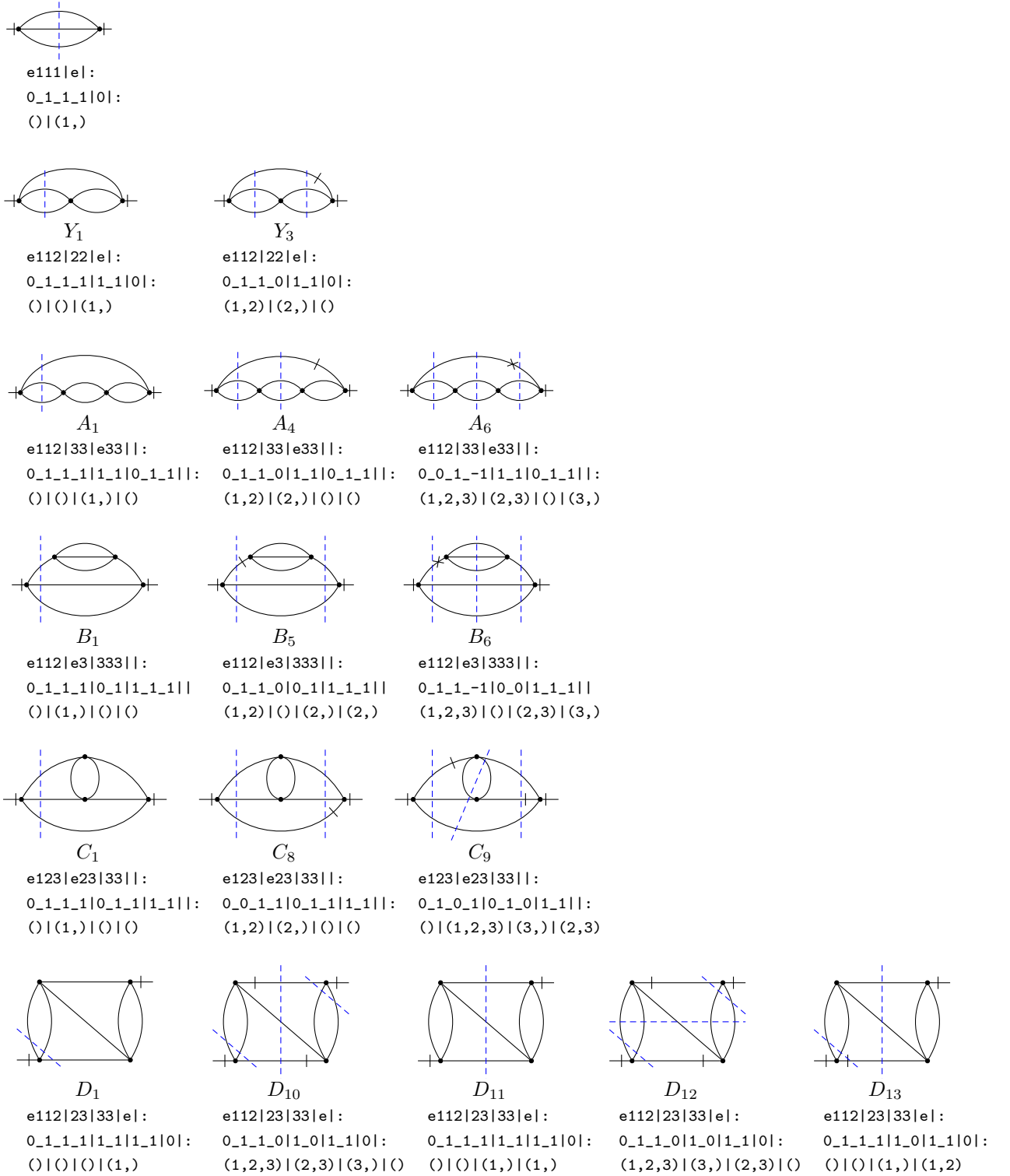

    \centering
    \begin{tabular}{c c c c c}
     \hspace{-10mm} \input{tikz_pics/dynamic_arbuz}   &  & & & \\

        {\footnotesize\makebox[60pt][l]{\texttt{\detokenize{e111|e|:}}}} & & & &
     \\
       {\footnotesize\makebox[60pt][l]{\texttt{\detokenize{0_1_1_1|0|:}}}} & & & &
     \\
       {\footnotesize\makebox[60pt][l]{\texttt{\detokenize{()|(1,)}}}} & & & &
     \\
     \\
      
     \hspace{-6mm} \input{tikz_pics/Y1_small}   &  \input{tikz_pics/Y3_small} & & & \\
      \hspace{-6mm} $Y_1 $ & $Y_3$ & & &\\

      {\footnotesize\makebox[60pt][l]{\texttt{\detokenize{e112|22|e|:}}}} &
       {\footnotesize\makebox[60pt][l]{\texttt{\detokenize{e112|22|e|:}}}} & & &
     \\
       {\footnotesize\makebox[60pt][l]{\texttt{\detokenize{0_1_1_1|1_1|0|:}}}} &
       {\footnotesize\makebox[60pt][l]{\texttt{\detokenize{0_1_1_0|1_1|0|:}}}} & & &
     \\
       {\footnotesize\makebox[60pt][l]{\texttt{\detokenize{()|()|(1,)}}}} &
       {\footnotesize\makebox[60pt][l]{\texttt{\detokenize{(1,2)|(2,)|()}}}} & & &
     \\
     \\     

      \input{tikz_pics/A1_small} & \input{tikz_pics/A4_small} & \input{tikz_pics/A6_small} & &\\
      $A_1$ & $A_4$ & $A_6$ & &\\ 

        {\footnotesize\makebox[60pt][l]{\texttt{\detokenize{e112|33|e33||:}}}} &
       {\footnotesize\makebox[60pt][l]{\texttt{\detokenize{e112|33|e33||:}}}} & {\footnotesize\makebox[60pt][l]{\texttt{\detokenize{e112|33|e33||:}}}} & &
     \\
       {\footnotesize\makebox[60pt][l]{\texttt{\detokenize{0_1_1_1|1_1|0_1_1||:}}}} &
       {\footnotesize\makebox[60pt][l]{\texttt{\detokenize{0_1_1_0|1_1|0_1_1||:}}}} & {\footnotesize\makebox[60pt][l]{\texttt{\detokenize{0_0_1_-1|1_1|0_1_1||:}}}} & &
     \\
       {\footnotesize\makebox[60pt][l]{\texttt{\detokenize{()|()|(1,)|()}}}} &
       {\footnotesize\makebox[60pt][l]{\texttt{\detokenize{(1,2)|(2,)|()|()}}}} & {\footnotesize\makebox[60pt][l]{\texttt{\detokenize{(1,2,3)|(2,3)|()|(3,)}}}} & &
     \\
     \\     
      
      \input{tikz_pics/B1_small} & \input{tikz_pics/B5_small} & \input{tikz_pics/B6_small} & & \\
      $B_1$ & $B_5$ & $B_6$ & & \\ 

      {\footnotesize\makebox[60pt][l]{\texttt{\detokenize{e112|e3|333||:}}}} &
       {\footnotesize\makebox[60pt][l]{\texttt{\detokenize{e112|e3|333||:}}}} & {\footnotesize\makebox[60pt][l]{\texttt{\detokenize{e112|e3|333||:}}}} & &
     \\
       {\footnotesize\makebox[60pt][l]{\texttt{\detokenize{0_1_1_1|0_1|1_1_1||}}}} &
       {\footnotesize\makebox[60pt][l]{\texttt{\detokenize{0_1_1_0|0_1|1_1_1||}}}} & {\footnotesize\makebox[60pt][l]{\texttt{\detokenize{0_1_1_-1|0_0|1_1_1||}}}} & &
     \\
       {\footnotesize\makebox[60pt][l]{\texttt{\detokenize{()|(1,)|()|()}}}} &
       {\footnotesize\makebox[60pt][l]{\texttt{\detokenize{(1,2)|()|(2,)|(2,)}}}} & {\footnotesize\makebox[60pt][l]{\texttt{\detokenize{(1,2,3)|()|(2,3)|(3,)}}}} & &
     \\
     \\  
      
      \input{tikz_pics/C1} & \input{tikz_pics/C8} & \input{tikz_pics/C9} \\
      $C_1$ & $C_8$ & $C_9$ \\ 

       {\footnotesize\makebox[60pt][l]{\texttt{\detokenize{e123|e23|33||:}}}} &
       {\footnotesize\makebox[60pt][l]{\texttt{\detokenize{e123|e23|33||:}}}} & {\footnotesize\makebox[60pt][l]{\texttt{\detokenize{e123|e23|33||:}}}} & &
     \\
       {\footnotesize\makebox[60pt][l]{\texttt{\detokenize{0_1_1_1|0_1_1|1_1||:}}}} &
       {\footnotesize\makebox[60pt][l]{\texttt{\detokenize{0_0_1_1|0_1_1|1_1||:}}}} & {\footnotesize\makebox[60pt][l]{\texttt{\detokenize{0_1_0_1|0_1_0|1_1||:}}}} & &
     \\
       {\footnotesize\makebox[60pt][l]{\texttt{\detokenize{()|(1,)|()|()}}}} &
       {\footnotesize\makebox[60pt][l]{\texttt{\detokenize{(1,2)|(2,)|()|()}}}} & {\footnotesize\makebox[60pt][l]{\texttt{\detokenize{()|(1,2,3)|(3,)|(2,3)}}}} & &
     \\
     \\  
      
      \input{tikz_pics/D1}  & \input{tikz_pics/D10} & \hspace{2mm} \input{tikz_pics/D11} & \hspace{2mm} \input{tikz_pics/D12} & \hspace{5mm} \input{tikz_pics/D13} \\
      $D_1$ & $D_{10}$ &\hspace{2mm} $D_{11}$ &\hspace{2mm} $D_{12}$ & \hspace{5mm} $D_{13}$ \\

       {\footnotesize\makebox[60pt][l]{\texttt{\detokenize{e112|23|33|e|:}}}} &
       {\footnotesize\makebox[60pt][l]{\texttt{\detokenize{e112|23|33|e|:}}}} &
       \hspace{2mm} {\footnotesize\makebox[60pt][l]{\texttt{\detokenize{e112|23|33|e|:}}}} & \hspace{2mm} {\footnotesize\makebox[60pt][l]{\texttt{\detokenize{e112|23|33|e|:}}}}  & 
       \hspace{5mm} {\footnotesize\makebox[60pt][l]{\texttt{\detokenize{e112|23|33|e|:}}}}
     \\
       {\footnotesize\makebox[60pt][l]{\texttt{\detokenize{0_1_1_1|1_1|1_1|0|:}}}} &
       {\footnotesize\makebox[60pt][l]{\texttt{\detokenize{0_1_1_0|1_0|1_1|0|:}}}} & \hspace{2mm} {\footnotesize\makebox[60pt][l]{\texttt{\detokenize{0_1_1_1|1_1|1_1|0|:}}}} & \hspace{2mm} {\footnotesize\makebox[60pt][l]{\texttt{\detokenize{0_1_1_0|1_0|1_1|0|:}}}} & 
       \hspace{5mm} {\footnotesize\makebox[60pt][l]{\texttt{\detokenize{0_1_1_1|1_0|1_1|0|:}}}}
     \\
       {\footnotesize\makebox[60pt][l]{\texttt{\detokenize{()|()|()|(1,)}}}} &
       {\footnotesize\makebox[60pt][l]{\texttt{\detokenize{(1,2,3)|(2,3)|(3,)|()}}}} & \hspace{2mm} {\footnotesize\makebox[60pt][l]{\texttt{\detokenize{()|()|(1,)|(1,)}}}} & \hspace{2mm} {\footnotesize\makebox[60pt][l]{\texttt{\detokenize{(1,2,3)|(3,)|(2,3)|()}}}} & 
       \hspace{5mm} {\footnotesize\makebox[60pt][l]{\texttt{\detokenize{()|()|(1,)|(1,2)}}}}
       \\
      
    \end{tabular}
    
    \caption{All two-, three- and four-loop diagrams obtained after the additional reduction that contribute to $\Gamma^R_{\psi'\psi'}$ at four-loop order. The notation of the diagrams is chosen to coincide with that of \cite{AIKV_4lSD17}.}
    \label{reduced_diags}
\end{figure}

The reduction of the diagrams of topologies B, C and D from \cite{AIKV_4lSD17} is performed in a similar manner. In Figure \ref{reduced_diags}, we present the set of all reduced diagrams, which contribute to $\Gamma^R_{\psi'\psi'}$ \eqref{gamma_expansion} at four-loop order, together with their corresponding Nickel-like indices \cite{N_nickel1,N_nickel2,GraphState14} (see Appendix~\ref{AppendNickel}). The application of the additional reduction allows us to decrease the number of dynamic subgraphs from 17 to only 1. It is a part of the diagram $B_6$ and cannot be removed with this technique. The procedure of the corresponding subtraction is explained in the next section. Moreover, the total number of diagrams is reduced from 17 to 14.

\section{Construction of renormalized integrals} \label{sec:5_Construction of renormalized integrals}

The main difficulty lies in the evaluation of integrals with subdivergencies. The application of the parametric integration imposes certain constraints on the choice of the renormalization scheme. First, it is well known that massive integrals are evaluated significantly less efficiently within this approach \cite{panzerPhD15}. For this reason, we choose $m=0$ in \eqref{gamma_expansion}, which leads to the consideration of massless one-scale $p$-integrals. Second, this method is applicable only to convergent integrals. Thus, one needs to construct finite renormalized integrals corresponding to the original divergent ones, calculate them, and then reconstruct pole parts which will contribute to the renormalization constant $Z_1$\eqref{MS}.
In this work, we follow the idea implemented in \cite{KP16}, which allows us to generate convergent integrands for integrals which initially have subdivergences. This is achieved with a BPHZ-like renormalization scheme introduced in \cite{Brown13}. The integrals obtained this way can already be evaluated analytically with the parametric integration.

After performing the modified reduction, only diagram $B_6$ from Fig.~\ref{reduced_diags} remains, containing the dynamic subgraph of type (c) from Fig.~\ref{subgraph_types}, which needs to be subtracted. There are also static subgraphs that require the corresponding subtractions. {As was mentioned in the previous section, the treatment of static subgraphs is much more easy than of dynamic ones, since one-scale static subgraphs admit a straightforward contraction procedure that reduces the number of necessary subtractions:} 
\begin{equation}
    \label{contraction}
    \vcenter{\hbox{\input{tikz_pics/static_loop_impulses}}} = (4 \pi)^{2-\varepsilon} \frac{\Gamma(1-\varepsilon)^2 \Gamma(\varepsilon)}{\Gamma(2-2\varepsilon)} \left(\frac{1}{p^2}\right)^\varepsilon =(4 \pi)^{2-\varepsilon} \frac{\Gamma(1-\varepsilon)^2 \Gamma(\varepsilon)}{\Gamma(2-2\varepsilon)} \cdot \input{tikz_pics/line5},
\end{equation}
where $\varepsilon$ is the power of the propagator in momentum space. It explicitly extracts the pole in $\varepsilon$, thereby removing the necessity of performing subtractions. The only resulting static subgraph, which cannot be eliminated this way is a triangle in $D_1$. 
\begin{figure}[H]
    \centering
    \setlength{\tabcolsep}{24pt}
    \begin{tabular}{c c}
       \input{tikz_pics/triangle1}  &  \input{tikz_pics/dynamic_arbuz1}\\
       $S_1$  & $S_2$
    \end{tabular}
    \caption{Two subgraphs in the set of the four-loop reduced diagrams, requiring separate subtractions.}
    \label{subgraphs for subtraction}
\end{figure}

\noindent The renormalized convergent integral can be constructed in the form
\begin{equation}
    \label{renorm_int}
    G^R = \partial_p \mathcal{R'} G \bigg|_{p=\mu},
\end{equation}
where $G$ is a considered diagram, $p$ is an external momentum. Since the Green function \eqref{gamma_expansion} depends only on the dimensionless ratio $\mu/p$, from now on we set $p=1$ by measuring the momentum $p$ in $\mu$ units. Differentiating with respect to $p$ eliminates the superficial divergence and $\mathcal{R'}$ is Bogolubov's operation, which recursively subtracts all subgraph divergencies. It is defined with Zimmermann's forest formula
\begin{equation}
    \label{R'}
    \mathcal{R'} G = \prod_{i} (1-K_i)G\,\,,
\end{equation}
where the product runs over all divergent subgraphs of the diagram $G$. Its explicit form is determined by the subtraction operation $\mathcal{K}$, which generally depends on the choice of the renormalization scheme. In our situation, the latter must allow us to perform the subtraction under the integral sign, resulting in the overall convergent integrand. Let us note that, despite the fact that we use the MS-scheme to determine the renormalization constant $Z_1$ \eqref{MS}, it is not suitable in this case, since it requires an explicit subtraction of pole parts. Instead, we use the already mentioned BPHZ approach, where counterterms are defined by the expansion around the renormalization point $p=1,\omega=0$ and can be subtracted at the integrand level. However, this scheme cannot be used in its original form due to the fact that condition $p=1$ is meaningful only for 2-point one-scale subgraphs. To make it applicable for the 3-point triangle in Fig.~\ref{subgraphs for subtraction}, we utilize the infrared rearrangement (IRR) technique \cite{V_irr8,chetyrkin1980}, which states that in logarithmically divergent diagrams under $\mathcal{K}\mathcal{R'}$-operation, it is possible to nullify some external momenta in an appropriate way to obtain a $p$-integral without introducing infrared divergences. When applied to the triangle subgraph, it yields 
\begin{equation}
    \label{triangle_IRR}
    \mathcal{K} \mathcal{R'}  \vcenter{\hbox{\input{tikz_pics/triangle1_impulses}}}  = \,\, \mathcal{K} \mathcal{R'} \vcenter{\hbox{\input{tikz_pics/triangle2_impulses}}} \: =  \: \: \vcenter{\hbox{\input{tikz_pics/triangle2_impulses}}} \left.\vphantom{\rule{0pt}{30pt}}\right|_{p_1=1}\,\,.
\end{equation}
\noindent As a result, the explicit definition of the $\mathcal{K}$ operation for the only subgraphs appearing in the four-loop approximation, $S_1$ and $S_2$ from Fig.~\ref{subgraphs for subtraction}, is the following:
\begin{equation}
\mathcal{K} G := 
\begin{cases}
\left. G_{\text{IRR}} \right|_{p = 1} & \text{if } G \text{ is static } S_1, \\
\left. G \right|_{p = 1, \omega = 0} & \text{if } G \text{ is dynamic } S_2.
\end{cases}
\end{equation}

Let us now discuss in detail the renormalization procedure for diagrams $B_6$ and $D_1$. According to \eqref{renorm_int}, the renormalized integral for $B_6$ is given by the following expression:
\begin{equation}
    \label{renorm_B6}
    \partial_p \mathcal{R'} \vcenter{\hbox{\input{tikz_pics/B6_R_}}} \left.\vphantom{\rule{0pt}{20pt}}\right|_{p=1} = \partial_p \vcenter{\hbox{\input{tikz_pics/B6_R_}}} \left.\vphantom{\rule{0pt}{20pt}}\right|_{p=1} - \vcenter{\hbox{\input{tikz_pics/dynamic_arbuz1_impulses}}} \left.\vphantom{\rule{0pt}{20pt}}\right|_{k=1} \times \,\, \partial_p \vcenter{\hbox{\input{tikz_pics/B6_R_2}}} \left.\vphantom{\rule{0pt}{20pt}}\right|_{p=1},
\end{equation}
where $k$ is the momentum flowing through the subgraph. Due to the trivial momentum dependence, differentiating with respect to $p$ only results in multiplication by the factor ${-2l\varepsilon}$, where l is the number of loops. Now, the original unrenormalized diagram can be recovered at $p=1$:
\begin{equation}
\label{unrenorm_B6}
\vcenter{\hbox{\input{tikz_pics/B6_R_}}} \left.\vphantom{\rule{0pt}{20pt}}\right|_{p=1} = - \frac{1}{8\varepsilon} \partial_p \mathcal{R'} \vcenter{\hbox{\input{tikz_pics/B6_R_}}} \left.\vphantom{\rule{0pt}{20pt}}\right|_{p=1} + \frac{1}{2} \vcenter{\hbox{\input{tikz_pics/dynamic_arbuz1_impulses}}} \left.\vphantom{\rule{0pt}{20pt}}\right|_{k=1} \times \,\, \vcenter{\hbox{\input{tikz_pics/B6_R_2}}}\left.\vphantom{\rule{0pt}{20pt}}\right|_{p=1}.
\end{equation}
The first term on the right side, being the convergent integral, can be evaluated using the hyperlogarithm method. The diagrams in the second term are free of subdivergencies and diverge only superficially (only have pole of the first order), so they are suitable for calculation as well. Combining all together, we obtain two pole terms for the $B_6$ diagram which contribute to $A^{(4)}$ in \eqref{gamma_expansion}. Generally, in the case of nested subgraphs appearing in the counterterms, they are processed according to the forest formula \eqref{R'}. 

The renormalization of diagram $D_1$ is carried out in an analogous manner. We begin by contracting the static loop subgraph with \eqref{contraction}, and then perform the subtraction on the infrared rearranged triangle subgraph \eqref{triangle_IRR}:
\begin{align}
    \label{renorm_D1}
    &\partial_p \mathcal{R'} \vcenter{\hbox{\input{tikz_pics/D1_small}}} \left.\vphantom{\rule{0pt}{30pt}}\right|_{p=1} 
    = (4 \pi)^{2-\varepsilon} \frac{\Gamma(1-\varepsilon)^2 \Gamma(\varepsilon)}{\Gamma(2-2\varepsilon)} 
    \,\, \partial_p \mathcal{R'} \vcenter{\hbox{\input{tikz_pics/D1_line}}} \left.\vphantom{\rule{0pt}{30pt}}\right|_{p=1} =
    \nonumber \\
    &= (4 \pi)^{2-\varepsilon} \frac{\Gamma(1-\varepsilon)^2 \Gamma(\varepsilon)}{\Gamma(2-2\varepsilon)} 
    \,\, \left[ \partial_p \vcenter{\hbox{\input{tikz_pics/D1_line}}} \left.\vphantom{\rule{0pt}{30pt}}\right|_{p=1} 
    - \vcenter{\hbox{\input{tikz_pics/triangle2_impulses_small}}} \left.\vphantom{\rule{0pt}{30pt}}\right|_{k=1} 
    \times \,\, \partial_p \vcenter{\hbox{\input{tikz_pics/dynamic_arbuz1_impulses2}}}\left.\vphantom{\rule{0pt}{25pt}}\right|_{p=1} \right].
\end{align}
The unrenormalized diagram is given by
\begin{equation}
\label{unrenorm_D1}
\vcenter{\hbox{\input{tikz_pics/D1_line}}} \left.\vphantom{\rule{0pt}{30pt}}\right|_{p=1} = - \frac{1}{8\varepsilon} \partial_p \mathcal{R'} \vcenter{\hbox{\input{tikz_pics/D1_line}}}\left.\vphantom{\rule{0pt}{30pt}}\right|_{p=1} + \frac{1}{2} \vcenter{\hbox{\input{tikz_pics/triangle2_impulses_small}}} \left.\vphantom{\rule{0pt}{30pt}}\right|_{k=1} \times \,\, \vcenter{\hbox{\input{tikz_pics/dynamic_arbuz1_impulses2}}} \left.\vphantom{\rule{0pt}{25pt}}\right|_{p=1}.
\end{equation}

\section{Parametric integration of dynamic diagrams}  \label{sec:6_Parametric integration of dynamic diagrams}

Parametric integration is based on the fact that a wide class of Feynman diagrams can be expressed in terms of Goncharov's polylogarithms, defined in terms of iterated integrals
\begin{equation}
G(w_1, \dots, w_n; x) = \int_0^x \frac{dt}{t - w_1} G(w_2, \dots, w_n; t).
\end{equation}
The recursion terminates at \( G(; x) \equiv 1 \). When all the indices vanish, the definition is extended as 
\begin{equation}
G(\underbrace{0, \ldots, 0}_{n}; x) \equiv \frac{\ln^n(x)}{n!}.
\end{equation}
The number of indices \( n \) is known as the weight of the GPL. 

Multiple Zeta Values (MZVs), defined as 
\begin{equation}
\zeta_{m_1,\ldots,m_n} \equiv \sum_{0<k_1<\cdots<k_n}^\infty \frac{1}{k_1^{m_1}\cdots k_n^{m_n}}\,,\,\,\,\,\,\,m_1,\dots,m_n \in \mathbb{N}\,,\,\,\,\,\,\,\, m_n>1\,,
\end{equation}
constitute a special case of GPLs:
\begin{equation}
\zeta_{m_1,\ldots,m_n} = (-1)^n G\left( \underbrace{0,\ldots,0}_{m_n-1}, 1, \underbrace{0,\ldots,0}_{m_{n-1}-1}, 1,\ldots, \underbrace{0,\ldots,0}_{m_1-1}, 1; 1 \right).
\end{equation}

GPLs satisfy a large number of relations, some of which follow from the shuffle rule \cite{R_shuffle58} for the product of two GPLs
\begin{equation}
G(a_1, \ldots, a_n; x)G(b_1, \ldots, b_m; x) = \sum_{c \in a \,\shuffle\, b} G(c_1, \ldots, c_{n+m}; x),
\end{equation}
where $a \,\shuffle \, b$ denotes the shuffles of the lists $a = \{a_1, \ldots, a_n\}$ and $b = \{b_1, \ldots, b_m\}$,i.e., all permutations that preserve the relative order of elements within each list. The MZVs inherit the same shuffle structure. 

The evaluation of renormalized integrals was carried out using \textit{HyperInt} -- a \textit{Maple}-based \cite{maple} implementation of the parametric integration algorithm \cite{Panzer15}. This method relies on a specific property of the integrand known as linear reducibility, which ensures that the integrand can be integrated sequentially in terms of GPLs at each step. This property may depend on the chosen order of integration. In \cite{Brown08}, a special method called \textit{polynomial reduction} (included in \textit{HyperInt}) was introduced, which allows one to verify the linear reducibility of a given integral in advance, prior to performing the integration itself. If an integral turns out to be linearly irreducible, there are a few possible strategies. The most straightforward one is to look for an integration order that satisfies this property. If none exists, another option is to perform a change of integration variables that restores linear reducibility. Such transformations can be difficult to identify, as they must be applied at a specific integration step and often take a non-trivial form. Some integrals cannot be expressed solely in terms of GPLs -- for instance, the massive two-loop sunset diagram in the $\phi^4$ model -- and require an extension of the functional class. Examples include elliptic polylogarithms \cite{EllipHpl18,EllipHpl19}, polylogarithms on K3 surfaces and more complex geometries \cite{BY_K318,D_K325}. 

Recently, considerable effort has been devoted to understanding the underlying reasons why certain Feynman diagrams are linearly reducible, while others are not. Since this property is directly related to the diagram's topology, a key question in the context of applying this method to critical dynamics is how the presence of time dependence — and consequently, dynamic propagators and time cuts \eqref{3l_tvs} — affects linear reducibility in comparison to the static case, such as the $\phi^4$ model. Answering this question would make it possible to assess the prospects of applying the parametric integration method to multiloop analytic calculations in other models of critical dynamics.

The first step in the parametric integration is to express the initial momentum integral in Feynman representation. A product of quadratic in momenta factors $D_i$, corresponding to propagators and cuts, can be re-expressed by introducing additional integrations over Feynman parameters $u_i$ 
\begin{equation}
\label{feym_param_1}
D_1^{-\lambda_1}...D_n^{-\lambda_n} = \frac{\Gamma(\sum \lambda_i)}{\prod \Gamma(\lambda_i)} \int_0^1 ... \int_0^1 du_1...du_n \frac{\delta(\sum u_i - 1) \prod u_i^{\lambda_i-1}}{\left[ \sum D_i u_i \right]^{\sum \lambda_i}}  
\end{equation}
Subsequently, the integrals over momenta are evaluated by the following formula:
\begin{equation}
\label{feym_param_2}
(2 \pi)^{-d l} \int ... \int \frac{d\mathbf{k_1}...d\mathbf{k_l}}{\left[ M_{is} k_i k_s + 2a_i k_i +c \right]^{\alpha}} = \frac{(4 \pi)^{-dl/2} \Gamma(\alpha - dl/2) (\det M)^{-d/2}}{\Gamma(\alpha)\left[ c - (M^{-1})_{is} a_i a_s \right]^{\alpha-dl/2}}  ,  
\end{equation}
where $M_{is}$ is the coefficient matrix and $a_i$ is the linear term vector of the quadratic form in the denominator of \eqref{feym_param_1}. The $\Gamma$-function in \eqref{feym_param_2} contains a pole due to the superficial divergence, while the subgraph divergences remain in the new parametric integral. 

All reduced diagrams in Fig. \ref{reduced_diags} were processed using \textit{HyperInt}. After their renormalization, described in Sec. \ref{sec:5_Construction of renormalized integrals}, they can be safely expanded in $\varepsilon$ to the required order for constructing $Z_1$ \eqref{MS} in the four-loop approximation - each coefficient is guaranteed to be finite. It appears that all diagrams except $C_9$ are linear reducible from the outset with a proper choice of integration order. Let us now discuss the computation of diagram $C_9$ in detail, as it illustrates specific challenges that are likely characteristic of dynamic diagrams in general.

The initial integrand is given by
\begin{align}
  & I(\mathbf{p}) = \vcenter{\hbox{\input{tikz_pics/C9_numbers}}}
    = \left[\frac{S_d}{(2\pi)^d}\right]^{-l} \int d\mathbf{k}\, d\mathbf{q}\, d\mathbf{\ell}\, d\mathbf{s} \, I^{(s)}(\mathbf{k},\mathbf{q},\mathbf{l},\mathbf{s},\mathbf{p})\,I^{(d)}(\mathbf{k},\mathbf{q},\mathbf{l},\mathbf{s},\mathbf{p})\\ 
    & I^{(s)} = \frac{1}{(k+p)^2 \, q^2 \, (k+\ell)^2 \, (q+s)^2 \, (\ell+s)^2}\,, \nonumber \\
    &I^{(d)} = \frac{1}{\left[ (k+p)^2 + (k+q)^2 + q^2 \right] \, \left[ (k+p)^2 + \ell^2 + (k+\ell)^2 \right] \, \left[ (k+p)^2 + q^2 + (q+s)^2 + (\ell+s)^2 + (k+\ell)^2 \right]} \, , \nonumber
\end{align}
where $I^{(s)}$ corresponds to the combination of static propagators $\langle\psi \psi\rangle_{st}$ and {$I^{(d)}$ corresponds to the result of integration over time.} The integrand is then transformed using Feynman parametrization \eqref{feym_param_1},\eqref{feym_param_2} and expanded in $\varepsilon$. It is known that the function $\delta(\sum u_i-1)$ in \eqref{feym_param_1} can be removed by extending the integration limits from $(0,1)$ to $(0,\infty)$ and fixing an arbitrary chosen Feynman parameter $u_i$ to $1$ \cite{ChengWu}. The topology of the diagram indicates that the middle time cut, labeled  $8$, would introduce a cubic nonlinearity in the corresponding parameter $u_8$, while the two other cuts yield quadratic nonlinearities in $u_6$ and $u_7$ (see Appendix~\ref{AppendA} for details). Therefore, it is natural to set $u_8=1$, as well as $p=1$, according to \eqref{renorm_int}. Since the diagram has four loops and contains no subdivergences, we are interested only in the coefficient of $\varepsilon^{-1}$, which corresponding integrand takes the following form:

\vspace{3 cm}
\begin{align}
\smash{
\begin{aligned}
&F(u_1,u_2,u_3,u_4,u_5,u_6,u_7) = \\
\frac{1}{64}&[ u_1 u_2 u_3 u_4 + u_1 u_2 u_3 u_5 + u_1 u_2 u_4 u_5 + 2 u_1 u_2 u_4 u_7 + 2 u_1 u_2 u_5 u_7 + u_1 u_3 u_4 u_5 + 2 u_1 u_3 u_4 u_6 + 2 u_1 u_3 u_5 u_6 \\
& + 2 u_1 u_4 u_5 u_6 + 2 u_1 u_4 u_5 u_7 + 4 u_1 u_4 u_6 u_7 + 4 u_1 u_5 u_6 u_7 + u_2 u_3 u_4 u_5 + 2 u_2 u_3 u_4 u_6 + 2 u_2 u_3 u_4 u_7 + 2 u_2 u_3 u_5 u_6 \\
&  + 2 u_2 u_3 u_5 u_7 + 2 u_2 u_4 u_5 u_6 + 2 u_2 u_4 u_5 u_7 + 4 u_2 u_4 u_6 u_7 + 3 u_2 u_4 u_7^2 + 4 u_2 u_5 u_6 u_7  + 3 u_2 u_5 u_7^2 + 2 u_3 u_4 u_5 u_6 \\
& + 2 u_3 u_4 u_5 u_7 + 3 u_3 u_4 u_6^2 + 4 u_3 u_4 u_6 u_7 + 3 u_3 u_5 u_6^2 + 4 u_3 u_5 u_6 u_7 + 3 u_4 u_5 u_6^2 + 6 u_4 u_5 u_6 u_7 + 3 u_4 u_5 u_7^2 \\
&  + 6 u_4 u_6^2 u_7 + 6 u_4 u_6 u_7^2 + 6 u_5 u_6^2 u_7 + 6 u_5 u_6 u_7^2 + + 2 u_1 u_2 u_3 + 2 u_1 u_2 u_4 + 2 u_1 u_2 u_5 + 4 u_1 u_2 u_7+ 2 u_1 u_2 u_3\\
&  + 2 u_1 u_2 u_4 + 2 u_1 u_2 u_5 + 4 u_1 u_2 u_7 + 2 u_1 u_3 u_4 + 2 u_1 u_3 u_5 + 4 u_1 u_3 u_6 + 2 u_1 u_4 u_5 + 4 u_1 u_4 u_6 + 4 u_1 u_4 u_7 \\
&  + 4 u_1 u_5 u_6 + 4 u_1 u_5 u_7 + 8 u_1 u_6 u_7 + 2 u_2 u_3 u_4 + 2 u_2 u_3 u_5 + 4 u_2 u_3 u_6  + 4 u_2 u_3 u_7 + 2 u_2 u_4 u_5 + 4 u_2 u_4 u_6\\
& + 6 u_2 u_4 u_7 + 4 u_2 u_5 u_6 + 6 u_2 u_5 u_7 + 8 u_2 u_6 u_7 + 6 u_2 u_7^2 + 2 u_3 u_4 u_5 + 6 u_3 u_4 u_6  + 4 u_3 u_4 u_7  + 6 u_3 u_5 u_6\\
& + 4 u_3 u_5 u_7 + 6 u_3 u_6^2 + 8 u_3 u_6 u_7 + 6 u_4 u_5 u_6  + 6 u_4 u_5 u_7 + 6 u_4 u_6^2 + 18 u_4 u_6 u_7 + 6 u_4 u_7^2  + 6 u_5 u_6^2 + 6 u_5 u_7^2\\
&   + 12 u_6^2 u_7  + 12 u_6 u_7^2 + 3 u_1 u_2 + 3 u_1 u_3 + 3 u_1 u_4 + 3 u_1 u_5 + 6 u_1 u_6 + 6 u_1 u_7 + 3 u_2 u_3 + 3 u_2 u_4 + 3 u_2 u_5\\
& + 6 u_2 u_6 + 10 u_2 u_7 + 3 u_3 u_4 + 3 u_3 u_5 + 10 u_3 u_6 + 6 u_3 u_7 + 3 u_4 u_5 + 10 u_4 u_6 + 10 u_4 u_7 + 10 u_5 u_6 \\
& + 10 u_5 u_7 + 9 u_6^2 + 30 u_6 u_7 + 9 u_7^2 + 4 u_1 + 4 u_2 + 4 u_3 + 4 u_4 + 4 u_5 + 14 u_6 + 14 u_7 + 5]^{-2}
\raisetag{-23ex}
\end{aligned}
}
\label{C9_full_int}
\end{align}

Performing the polynomial reduction via \textit{HyperInt} allows one to determine whether an integral is linear reducible and to identify the upper boundary of a set of irreducible polynomials that may occur after integrating out certain variables. It turns out that the integrand \eqref{C9_full_int} is linear reducible in all static variables $u_1...u_5$, with the following set of potential nonlinear polynomials
\begin{equation}
\begin{aligned}
    L_{\{u_1,u_2,u_3,u_4,u_5\}} = & \left\{  u_7+u_6^2+\frac{5}{3} u_6+\frac{4}{3}u_6u_7+\frac{2}{3}\,,\,\, \frac{5}{4}u_7+\frac{3}{4} u_7^2 + \frac{3}{4}u_6 + u_6 u_7 + \frac{1}{2}\,, \right. \\
    & \left. \,\,\,\, u_6^2 + \frac{4}{3}u_6 u_7 + \frac{4}{3} u_6 + \frac{2}{3} u_7 + \frac{1}{3}\,,\,\, u_6 u_7 + \frac{3}{4} u_7^2 + \frac{1}{2} u_6 + u_7 + \frac{1}{4} \right\} \,, \\
\label{C9_init_poly_set}
\end{aligned}
\end{equation}

\vspace{-0.5cm}
\noindent which remains the same for all possible integration orders of $\{u_1,u_2,u_3,u_4,u_5\}$ - two quadratic polynomials in each of $u_6$ and $u_7$. However, these integration orders are not equivalent in terms of linear reducibility in the next integrations. We found it to be optimal to integrate over all of them directly and manually inspect the resulting sets of polynomials. It appears that while some orders indeed produce  the full set \eqref{C9_init_poly_set} predicted by the polynomial reduction, others result in a smaller set, containing only one quadratic polynomial in each of $u_6$ and $u_7$. For instance, the order $(u_2,u_1,u_3,u_5,u_4)$, which we will adopt from now on, yields
\begin{equation}
\begin{aligned}
    L_{\{u_2,u_1,u_3,u_5,u_4\}} =  \left\{ u_6^2+\frac{4}{3}u_6u_7+\frac{5}{3}u_6+u_7+\frac{2}{3}\,,\,\, \frac{3}{4}u_7^2 + u_6u_7+\frac{3}{4}u_6+\frac{5}{4}u_7+\frac{1}{2} \right\} \,.
\label{C9_better_poly_set}
\end{aligned}
\end{equation}
The smaller set of quadratic polynomials can be viewed as evidence of a more favorable integration order\footnote{We would like to note that the \textit{HyperInt}'s \textit{suggestIntegrationOrder} command actually recommends the sequence $(u_4,u_5,u_2,u_1,u_3)$, which produces the suboptimal larger set \eqref{C9_init_poly_set}.}.

A standard approach to handling irreducible quadratic polynomials, such as those in \eqref{C9_better_poly_set}, is to perform an appropriate change of variables that factorizes them. The \textit{RationalizeRoots} package \cite{RatioRoots} is particularly useful to automate this step. In our case, the polynomial $ u_6^2+4u_6u_7/3+5u_6/3+u_7+2/3$ is factorized by substitution $u_7=(2\tilde{u}_7+4)/(\tilde{u}^2_7-16)$, enabling us to perform the integration over $u_6$. Formally, the presence of nonlinear polynomials in the integrand at the last integration over $u_7$ does not pose any problems, since they can be factorized by introducing radicals that would enter into the alphabet of the resulting GPLs. However, in our case, this would lead to the appearance of the irratinonal letter $\sqrt{301}$ and would yield an excessively long expression.

We propose the application of an alternative approach to the treatment of nonlinear polynomials, briefly outlined in \cite{Brown09}, which appears to be better suited for the diagram considered. Moreover, it may prove beneficial in cases where no appropriate change of variable can be found to factorize complicated sets of polynomials. 
The main idea is to split the terms of the integrand $F$ into distinct combinations $\tilde{f_1}$ and $\tilde{f_2}$, each of which is linearly reducible with respect to a specific integration order. Then, taking into account the finiteness of the original integral, one can construct appropriate counterterm $\delta f$ to ensure the finiteness of integrals $f_1 = \tilde{f_1} + \delta f$ and $f_2 = \tilde{f_2} -  \delta f$. This allows for the construction of linearly reducible parts $f_1$ and $f_2$, whose integrals are convergent. We refer to them as \textit{''streams''}. For example, if $f_1$ is integrable in the order $[u_6, u_7]$ and $f_2$ in $[u_7, u_6]$, and all corresponding integrals converge, then the following relation holds:

\begin{eqnarray}
    \label{int_decomp}
    \int_0^{\infty}du_6 \int_0^{\infty} du_7 \int_0^{\infty}d\{u\}_{st} \,\, F(\{u\}_{st},u_6,u_7) = \int_0^{\infty}du_7 \int_0^{\infty} du_6 \int_0^{\infty}d\{u\}_{st} \,\, f_1(\{u\}_{st},u_6,u_7) \,\, + \\
    +\,\, \int_0^{\infty}du_6 \int_0^{\infty} du_7 \int_0^{\infty}d\{u\}_{st} \,\, f_2(\{u\}_{st},u_6,u_7) \,\,,\nonumber
\end{eqnarray}
where $\{u\}_{st}$ denotes the set of static variables. 

The decomposition procedure can be performed after any number of integrations over the static variables. It would seem desirable to choose the integration order such that this number is maximized. It seems natural to perform all possible integrations over linear reducible variables $\{u\}_{st}$ and then split the resulting integrand into streams. However, in our case, it turned out to be more convenient to integrate over two variables first, namely $[u_2, u_1]$. At this stage, the resulting expression contains fewer terms compared to the case where integration is carried out over all five static variables, which significantly simplifies the analysis of linear reducibility and the subsequent grouping into \textit{streams}. In this formulation, one stream is integrable in the sequence $[u_3, u_5, u_4, u_6, u_7]$, while the other proceeds in the sequence $[u_3, u_5, u_4, u_7, u_6]$. Each stream can be integrated independently over its respective variable order. We note that the procedure of separating terms into streams is applicable only when no terms contain quadratic polynomials simultaneously in both variables, which fortunately holds in this case.

To decompose this expression into \textit{streams}, one can analyze the linear reducibility of each term in $F$ and group the terms accordingly. Naturally, some terms may be linearly reducible in both variables; these can either be assigned to any stream or isolated into a separate one. Fortunately, in our case, the two terms produced by \textit{HyperInt} after the first two integrations over $[u_2,u_1]$ are linearly reducible in the corresponding integration orders and can be used as $\tilde{f_1}$ and $\tilde{f_2}$:

\begin{equation}
\label{stream1}
\begin{aligned}
     \tilde{f_1}  =  & - \frac{1}{64} \ln \Big[ ((3u_3 + 3u_5 + 6u_7 + 6)u_4 + (3u_3 + 6u_7 + 6)u_5 + 6u_3 + 12u_7 + 9)u_6^2 + ((2u_3 + 6u_7 + 6)u_5 + \\
     & 6u_7^2 + (4u_3 + 18)u_7 + 6u_3 + 10)u_4 + (6u_7^2 + (4u_3 + 18)u_7 + 6u_3 + 10)u_5 + 12u_7^2 + (8u_3 + 30)u_7 + \\ 
     & 10u_3 + 14)u_6 + (2(u_7 + 1)(3u_7/2 + u_3 + 3/2)u_5 + 6u_7^2 + (4u_3 + 10)u_7 + 3u_3 + 4)u_4 + (6u_7^2 + \\
     & (4u_3 + 10)u_7 + 3u_3 + 4)u_5 + 9u_7^2 + (6u_3 + 14)u_7 + 4u_3 + 5) / (((2u_3 + 2u_5 + 4u_7 + 4)u_4 + (2u_3 + \\
     & 4u_7 + 4)u_5 + 4u_3 + 8u_7 + 6)u_6 + ((u_3 + 2u_7 + 2)u_5 + 2u_3 + 4u_7 + 3)u_4 + (2u_3 + 4u_7 + 3)u_5 + 3u_3 \\
     & + 6u_7 + 4) \Big] \Big/ \Big[ \left( (u_3 + u_6 + u_7 + 1)u_5 + (u_3 + 2u_7 + 2)u_6 + u_3 + u_7 + 1 \right)u_4 + \left( (u_3 + 2u_7 + 2)u_6 + \right. \\
     & \left. u_3 + u_7 + 1 \right)u_5 + (2u_3 + 4u_7 + 3)u_6 + u_3 + u_7 + 1 \Big]^2
\end{aligned}
\end{equation}

\begin{equation}
\label{stream2}
\begin{aligned}
     \tilde{f_2}  =  & - \frac{1}{64} \ln \Big[ \big( \left( (u_3 + 2u_6 + 2u_7 + 2)u_5 + 3u_7^2 + (2u_3 + 4u_6 + 6)u_7 + (2u_6 + 2)u_3 + 4u_6 + 3 \right)u_4 + \\
    &  \left( 3u_7^2 + (2u_3 + 4u_6 + 6)u_7 + (2u_6 + 2)u_3 + 4u_6 + 3 \right)u_5 + 6u_7^2 + (4u_3 + 8u_6 + 10)u_7 + \\ 
    & (4u_6 + 3)u_3 + 6u_6 + 4 \big) \big/ \left( (u_3 + u_5 + 2u_7 + 2)u_4 + (u_3 + 2u_7 + 2)u_5 + 2u_3 + 4u_7 + 3 \right) \Big] \Big/ \\
    & \Big[ ((u_3+u_6+u_7+1)u_5+(u_3+2u_7+2)u_6+u_3+u_7+1)u_4+((u_3+2u_7+2)u_6+u_3+ \\
    &  u_7+1)u_5 +  (2u_3+4u_7+3)u_6+u_3+u_7+1 \Big]^2
\end{aligned}
\end{equation}

 The process of finding counterterms $\delta f$ to ensure the convergence of integrals on the right hand side of equation \eqref{int_decomp} can be automated by enabling the \textit{HyperInt} parameter \textit{hyper\_check\_divergences}, which stores all emerging divergences. In our case, the stream $\tilde{f_1}$ contains a divergence of the form $\ln(u_7=\infty)$ with coefficient
\begin{equation}
\begin{aligned}
    c_1(u_6) = & \frac{1}{64}\frac{1}{2u_6^2} \big[ (-4u_6+2) \, G(1/2, -1/2; u_6) + (4 u_6-2) \, G(1/2,-1/4;u_6)+(-4 u_6+4) \, G(1,-1/2;u_6) - \\
    &  4(1+\ln(2)-\ln(3)) (u_6+1/4)\,G(-1/4;u_6) + ((-4u_6-2)\ln(3)+8u_6+4)\,G(-1/2;u_6) + \\
    & 4 (u_6-1/2) (\ln(2)-\ln(3)) \, G(1/2;u_6)+(4 u_6-4)\,G(1,-1;u_6) + 4 (-1+\ln(2)) (u_6+1)\,G(-1;u_6) - \\
    & 4\,G(1;u_6)(\ln(2)-\ln(3))(u_6-1) \big]
\end{aligned}
\end{equation}
and a divergence of the form $\ln(u_7=\infty)^2$ with coefficient
\begin{equation}
    c_2 (u_6) = \frac{1}{64} \frac{1}{4u_6^2} \big[ (4+8u_6)\,G(-1/2;u_6)+(-1-4u_6)\,G(-1/4;u_6)-(4(u_6+1))\,G(-1;u_6) \big]\,\,.
\end{equation}
Hence, the appropriate counterterm has the form
\begin{equation}
    \label{counterterm_final}
    \delta f= \left(  \frac{1}{u_7+1} \, c_1(u_6) + \left(-\frac{2 \ln(u_7+1)}{u_7+1}\right) c_2(u_6) \right) \, g(u_3) \, g(u_5) \, g(u_4)\, , \,\,\,\,\,\,\,\,\,\, g(x) \equiv \frac{2}{\pi (1+x^2)}\, ,
\end{equation}
where the factors in front of $c_1$ and $c_2$ produce the required logarithmic divergences upon integration, and the auxiliary function $g(x)$ leaves the counterterm intact when integrating over $(u_3,u_5, u_4)$. Thus, we choose it to be a polynomial (integrable with \textit{HyperInt}) that evaluates to 1 after the corresponding integration.

Adding and subtracting the counterterm \eqref{counterterm_final} from the streams \eqref{stream1}, \eqref{stream2}, we obtain the finite integrands, which can be safely integrated in their corresponding sequences. The final integration in both streams can be carried out by factoring the resulting nonlinear polynomials with the required radicals. Compared to the change-of-variable method, stream-based integration leads to a considerably cleaner GPL alphabet; however, the resulting expression remains extremely cumbersome in contrast to those of other diagrams. 

Given the length of the analytic expressions obtained for all diagrams, they are provided in the Supplementary Material together with their numeric evaluation of  up to 200 digits performed with \textit{Ginac} \cite{Ginac1,Ginac2}. In Table \ref{tab:diag_res}, we present their values only up to 15 digits for illustration purposes, along with the corresponding factors $S$ - which include both the symmetry coefficient and additional contributions from the reduction procedure -  and weight factors $f(n)$. The latter, all equal to $1$ with $n=1$, allow us to obtain results for the theory with an arbitrary number of components $n$ of the order parameter. Note that  factors $S$ already contain $1/2$ from the definition of the Green function considered $\Gamma_{\psi' \psi'} = \langle \psi' \psi' \rangle_{1-\text{irr}} / (2 \lambda)$. For verification, the diagrams were also computed using the numerical sector decomposition method; the corresponding results are listed in Table \ref{tab:diag_res} as well and agree with the analytic values within the error margins.

\begin{table}[H]
    \centering
    \begin{tabular}{|c|c|c|c|c|c|}
        \hline
        № &   $S$  & $f(n)$ & \multicolumn{3}{c|}{\textbf{Number of loops}} \\ \hline
         \multicolumn{3}{|c|}{}     & \multicolumn{3}{c|}{\textbf{2 loop}\tablefootnote{The two-loop diagram has two identical time versions, which is taken into account in $S$.}} \\ \cline{4-6}
         \multicolumn{3}{|c|}{}     & $\varepsilon^{-1}$ & $\varepsilon^{0}$ & $\varepsilon^{1}$ \\ \hline
         & 1/6  &   $(2+n)/3$  & \makecell[l]{0.107880777169417 \\ 0.107880(2)} & \makecell[l]{0.250021247245667 \\ 0.250027(20)} & \makecell[l]{0.777492216471051 \\ 0.77754(5)} \\ \hline

          \multicolumn{3}{|c|}{}     & \multicolumn{3}{c|}{\textbf{3 loop}} \\ \cline{4-6}
          \multicolumn{3}{|c|}{}     & $\varepsilon^{-2}$ & $\varepsilon^{-1}$ & $\varepsilon^{0}$ \\ \hline
        $Y_1$ & 1/2    &  \multirow{2}{*}[-1.1ex]{$\displaystyle \frac{(8+n)(2+n)}{27}$}  & \makecell[l]{0.0359602590564726 \\ 0.0359593(15)} & \makecell[l]{0.152369545824030 \\ 0.152364(9)} & \makecell[l]{0.587196703355624 \\ 0.58719(4)} \\ \cline{1-2} \cline{4-6}
        $Y_3$ & -1/4   &   & \textbf{--} & \makecell[l]{0.0310353655538069 \\ 0.0310358(14)} & \makecell[l]{0.0907157790941646 \\ 0.090714(10)} \\ \hline

        \multicolumn{3}{|c|}{}    & \multicolumn{3}{c|}{\textbf{4 loop}} \\ \cline{4-6}
        \multicolumn{3}{|c|}{}   & $\varepsilon^{-3}$ & $\varepsilon^{-2}$ & $\varepsilon^{-1}$ \\
         \hline
        $A_1$ & 3/8   &  \multirow{2}{*}[-4ex]{$\displaystyle \frac{(n^2+6n+20)(2+n)}{81}$} & \makecell[l]{0.0134850971461771 \\ 0.013489(21) } & \makecell[l]{0.0830245034623135 \\ 0.08286(18)} & \makecell[l]{0.391383732451479 \\ 0.3922(11)} \\  \cline{1-2} \cline{4-6}
        $A_4$ & -3/8   &    & \textbf{--} & \makecell[l]{0.0116382620826775 \\ 0.011656(25) } & \makecell[l]{0.05328447601120 \\ 0.05324(19) } \\ \cline{1-2} \cline{4-6}
        $A_6$ & 1/8   &    & \textbf{--} & \textbf{--} & \makecell[l]{ 0.0100443580670449 \\ 0.01030(29)} \\ \hline
        $B_1$ & 1/12   &  \multirow{2}{*}[-4ex]{$\displaystyle \frac{(2+n)^2}{9}$} & \textbf{--} & \makecell[l]{ -0.0033712742865443 \\ -0.00337(4) } & \makecell[l]{ -0.0291843115819383 \\ -0.02906(26) } \\ 
        \cline{1-2} \cline{4-6}
        $B_5$ & 1/12   &    & \textbf{--} & \makecell[l]{-0.00112375809551477 \\ -0.0011255(27)} & \makecell[l]{-0.00881974186339340 \\ -0.008799(24)} \\ 
        \cline{1-2} \cline{4-6}
        $B_6$ & 1/12  &    & \textbf{--} & \makecell[l]{0.00193971034711292 \\ 0.001931(8)} & \makecell[l]{0.007822241167974 \\ 0.00781(5)} \\ \hline
        $C_1$ & 1/2   &  \multirow{2}{*}[-17.8ex]{$\displaystyle \frac{(22+5n)(2+n)}{81}$}  & \makecell[l]{0.0067425485730886 \\ 0.00674257(5) } & \makecell[l]{0.048254800304246 \\ 0.048256(4)} & \makecell[l]{0.231100647771624 \\ 0.231104(4)} \\
        \cline{1-2} \cline{4-6}
        $C_8$ & -1/4   &    & \textbf{--} & \makecell[l]{0.0116382620826767 \\ 0.011644(25) } & \makecell[l]{0.0445872837715371 \\ 0.04458(19) } \\ \cline{1-2} \cline{4-6}
        $C_9$ & 1/2    &    & \textbf{--} & \textbf{--} & \makecell[l]{0.00732642134804625 \\ 0.0073264206(25) } \\ \cline{1-2} \cline{4-6}
        $D_1$ & 1/2   &    & \makecell[l]{0.0067425485730886 \\ 0.00674257(5)} & \makecell[l]{0.048254800304246 \\ 0.048256(4) } & \makecell[l]{0.257431763676169 \\ 0.257447(27)} \\ \cline{1-2} \cline{4-6}
        $D_{10}$ & 1/4    &    & \textbf{--} & \textbf{--} & \makecell[l]{0.0100443580670446 \\ 0.010053(25)} \\ \cline{1-2} \cline{4-6}
        $D_{11}$ & 1/4    &    & \makecell[l]{0.0134850971461771 \\ 0.013483(23)} & \makecell[l]{0.0830245034623138 \\ 0.08292(18)} & \makecell[l]{0.376285778866138 \\ 0.3753(11)} \\ \cline{1-2} \cline{4-6}
        $D_{12}$ & 1/4   &    & \textbf{--} & \textbf{--} & \makecell[l]{0.005482183430494 \\ 0.005476(16)} \\ \cline{1-2} \cline{4-6}
        $D_{13}$ & -1/2  &     & \textbf{--} & \makecell[l]{0.0116382620826775 \\ 0.011634(25)} & \makecell[l]{0.053284476011194 \\ 0.05314(20)} \\
        \hline
        
    \end{tabular}
    \begin{otherlanguage}{english}
    \caption{Values of the coefficients in $\varepsilon$-expansion of the diagrams from Fig. \ref{reduced_diags}. $S$ denotes the prefactor arising from the symmetry factor and additional contributions from the diagram reduction procedure; $f(n)$ is a weight factor specific for the $n$-component theory. For each coefficient, the upper value corresponds to the numerical evaluation of the exact analytic expression evaluated with \textit{HyperInt}, while the lower value is obtained using the numerical sector decomposition method.}
    \label{tab:diag_res}
    \end{otherlanguage}
\end{table}

\section{Calculation of the dynamic exponent $z$}
\label{sec:7_Calculation of the dynamic exponent z}
Using the $\varepsilon$-expansions of the diagrams (Table \ref{tab:diag_res}), calculated up to the required order, the renormalization constant $Z_1$ \eqref{MS} can be found analytically up to the four loop. The explicit expressions for the coefficients $A_{ij}$ in \eqref{MS} are provided in Appendix~\ref{AppendB}. Here we only give the resulting expressions for $b_{i1}$, where $b_{41}$ is given numerically due to the large contribution of the $C_9$ diagram:
\begin{align}
\label{our_b31}
b_{21} &= -\frac{2+n}{48}\,\ln(4/3), \nonumber \\[6pt]
b_{31} &= \frac{(n+2)(n+8)}{2592}\Big[ 4\mathrm{Li}_2(-1/3)+3\ln(4/3)-\frac{17}{2}\ln^2(4/3) + \frac{1}{6}\pi^2 \Big] ,\nonumber \\[6pt]
b_{41} &= -0.00702883891635665-0.00484559952481972\,n-0.000609864256517009\,n^2+0.0000278628884018446\,n^3.
\end{align}
The analytic three-loop result $b_{31}$ was first obtained in \cite{AV_3l84} four decades ago, where the authors employed a different evaluation technique and managed to express the results in terms of dilogarithms. In Appendix~\ref{AppendC}, we provide a simplification of our initial result, expressed in terms of GPLs, which leads to  $b_{31}$ in \eqref{our_b31} and demonstrates that it is equivalent to the one reported in \cite{AV_3l84}.

The renormalization constant $Z_1$ corresponds to the RG-function $\gamma_1$
\begin{equation}
    \gamma_1(u)=\beta(u)\partial_u\ln Z_1\,,
\end{equation}
where $\beta(u)$ is the RG beta-function. In fact, because of the aforementioned connection between the higher poles to the simple pole, $\gamma_1$ can be obtained in a simpler way:
\begin{equation}
    \label{gamma1_simple}
    \gamma_1(u)=-2u\,\partial_u Z_1^{(1)}
\end{equation}

The exponent $z$ is expressed in terms of $\gamma_1^*\equiv\gamma_1(u_*)$, where $u_*$ is the fixed point, and the Fisher exponent $\eta$ by the relation
\begin{equation}
    z=2+\gamma_1^{*}-\eta\,.
\end{equation}
The expressions for $u_*$ and $\eta$ can be used from the theory $\phi^4$ with the corresponding precision \cite{KP17}:
\begin{equation}
    \label{u_star}
    u_*=\frac{12}{n+8}\varepsilon + \frac{72(3n+14)}{(n+8)^3}\varepsilon^2 + \frac{6}{(n+8)^5}\left[-33n^3+110 n^2+1760n+4544-96(n+8)(5n+22)\zeta(3)\right]\varepsilon^3 + \mathcal{O}(\varepsilon^4)
\end{equation}
\begin{align}
\label{eta_exp}
\eta =& \frac{2(n+2)}{(n+8)^{2}}\,\varepsilon^{2}
+ \frac{(n+2)\bigl(-n^{2}+56n+272\bigr)}{(n+8)^{4}}\,\varepsilon^{3} + \frac{(n+2)}{2\,(n+8)^{6}}
\Bigl[-5n^{4}-230n^{3}+1124n^{2}+17920n+\,\,\,\,\,\,\,\,\,\,\nonumber\\
&+46144-384\,(n+8)(5n+22)\,\zeta(3)\Bigr]\,\varepsilon^{4}+ \mathcal{O}(\varepsilon^4).
\end{align}
The $\varepsilon$-expansion for the dynamic exponent $z$ for the n-component order parameter in the four-loop approximation is given by
\begin{align}
    z(\varepsilon,n) =\:  2 \, + \, &\varepsilon^2\,\frac{2(n+2)}{(n+8)^2}\Big[6\ln(4/3)-1\Big] + \nonumber\\
& \varepsilon^3\,\frac{n+2}{(n+8)^2}\Big[ 1 - 12 \ln(4/3) + 34\ln^2(4/3) - \frac{2}{3}\pi^2 - 16 \, \mathrm{Li}_{2}(-1/3) + \frac{24(3n+14)}{(n+8)^2}\left( 6\ln(4/3)-1 \right) \Big]  + \nonumber\\ & 
\varepsilon^4\,\frac{n+2}{(n+8)^6}\Big[-11875.75943931894683515783928518 -3464.256191200665479956889326654\,n \nonumber\\ & 
-385.0912660888331502164857353582\,n^2-85.23257289375023905436261208910\,n^3 \nonumber\\ & 
-2.122118831205199575098764736384\,n^4\Big] + \mathcal{O}(\varepsilon^5) \,.
\end{align}
A partial verification of the analytic results obtained can be achieved by comparison with the $1/n$-expansion \cite{HHM_2l2}.  More details are provided in Appendix~\ref{AppendD}.

\section{Summary}
\label{sec:8_Summary}
In this work, we carried out a four-loop analytic RG calculation of the dynamic exponent $z$ in the $n$-component model A of critical dynamics. We implemented a modified diagram-reduction approach, based on the original method of \cite{AIKV_4lSD17}, which allowed us not only to reduce the number of dynamic diagrams, but also to eliminate several divergent dynamic subgraphs. From the reduced diagrams, we constructed the renormalized integrals, which enabled parametric integration using the \textit{Maple}-based package \textit{HyperInt} \cite{Panzer15} — a reliable method for analytic evaluation of a wide class of Feynman integrals \cite{Brown08}. 

The main requirement for this method is linear reducibility of the integral, which guarantees that the integrand can be expressed in terms of GPLs at every stage of integration. All integrals obtained met this condition, with the sole exception of diagram $C_9$. The standard procedure for treating linearly irreducible integrals is to find an appropriate change of variables that factorize nonlinear polynomials in the integrand \cite{RatioRoots}. For $C_9$ this was not feasible because integration over the static Feynman parameters produces the linearly non-reducible set of polynomials \eqref{C9_init_poly_set}, consisting of two pairs of quadratic polynomials in the remaining dynamic parameters. As a result, no change of variables can restore linear reducibility. Reordering the integration over the static parameters reduces the set to only two polynomials \eqref{C9_better_poly_set}, which in principle makes an appropriate variable transformation possible and allows one to complete the integration. However, this approach yields an extremely large final expression with a long polylogarithmic alphabet, clearly indicating that it is not an optimal strategy.

We proposed an alternative treatment for such integrals based on separating the integrand into \emph{streams}, each containing terms that are linearly reducible in the corresponding order of integration. By carefully accounting for the divergences that emerge from this separation, we obtained an analytic result expressed in terms of GPLs. To validate the analytic calculations, the diagrams were also evaluated numerically using the sector decomposition method. The results obtained are consistent with the numerical five-loop estimates of \cite{AEHIKKZ_5l_2_22} within the stated uncertainties.

Previously, RG studies of model A were carried out: analytically to three-loop order in \cite{AV_3l84} and numerically to five-loop order in \cite{AEHIKKZ_5l_2_22}. Therefore, our result does not improve the precision of the perturbative RG estimates for the exponent $z$. However, the main aim of this work is different: to test the feasibility of performing analytic multiloop calculations in models of critical dynamics.

The first such attempt was made in \cite{AEKTurb2024}, where the authors studied a model of stochastic turbulence in the limit $d\to\infty$ at four-loop order. Because that study considered this specific limit -- which drastically simplifies the diagrammatic structure -- it did not exhibit the complications characteristic of dynamic models. Model A is known to be the simplest in the Hohenberg-Halperin classification \cite{Halperin77}, making it an ideal testing ground for multiloop calculation strategies.

The nontrivial linear irreducibility of diagram $C_9$ appears to be a direct consequence of the presence of dynamic cut propagators. It is highly likely that at the next loop order many more diagrams will exhibit the same obstruction to linear reducibility. Similarly, more complicated dynamical models (indeed, essentially any other model) would almost certainly produce such problematic diagrams at orders below four. Thus, it is of great interest to determine whether the proposed \emph{separation into streams} method can serve as a general tool for evaluating complicated dynamical diagrams across different models.

The comparison with the $1/n$-expansion presented in Appendix~\ref{AppendD} provides a partial verification of the analytic results obtained. More importantly, it demonstrates that our results expressed in terms of GPLs can be substantially simplified. Generalized polylogarithms are known to satisfy a vast number of relations, which often allow for a drastic reduction of their combinations. A prominent example is the simplification of a 17-page result for the two-loop six-point function in $\mathcal{N}=4$ super Yang–Mills theory to only a few lines \cite{G_symbol10}. Such reductions are achieved by exploiting their algebraic structure, since GPLs form a Hopf algebra, which enables the use of the symbol and co-product approach \cite{D12_symbol,D19_symbol}.

In the present work, we do not attempt such a treatment of our GPL expressions, as they are sufficiently compact to be evaluated numerically with arbitrary precision in a reasonable time. In our case, such a simplification—although highly nontrivial in general—would provide only an aesthetic benefit. Nevertheless, it remains an interesting direction for future study, as it may shed light on which transcendentalities appear in dynamical QFT models and how they are related to those in static models.

\section*{Acknowledgments}
We are deeply grateful to E.~Panzer for his comments on hyperlogarithms and to A.~Kotikov for his advices on the calculation of the $C_9$ diagram. We would also like to thank E. Zerner-Käning for careful reading and editing. 
D.A.D., D.A.E. and M.V.K.  gratefully acknowledge the support of Foundation for the Advancement of Theoretical Physics "BASIS" through Grant 25-1-2-48-1.
The work of D.A.D. was performed at the Saint Petersburg Leonhard Euler International Mathematical Institute and supported by the Ministry of Science and Higher Education of the Russian Federation (agreement no. 075–15–2025–343).

\appendix

\refstepcounter{section}             
\phantomsection
\section*{Appendix \Alph{section}: Nickel notation for time versions}
\label{AppendNickel}
The indices in Fig.~\ref{reduced_diags} corresponding to each diagram are based on the Nickel notation \cite{N_nickel1, N_nickel2}, which assigns a unique string to an arbitrary diagram. This allows automated generation, identification, and manipulation of diagrams. Originally introduced for numbering static diagrams, it can be generalized to account for graphs with additional properties of edges and vertices.

Since we consider time versions of dynamic diagrams, the power of static propagators and the structure of time cuts must be uniquely encoded in the index. To achieve this, we use the framework from \cite{GraphState14}, which provides an index consisting of three parts, separated by '':''. The first part is the standard Nickel index defining the topology of the diagram. The second part specifies the power of static propagator corresponding to each edge: ''1'' denotes $\langle \psi \psi\rangle$ line, ''0'' is for $\langle \psi \psi'\rangle$ line, and ''-1'' -- line with an inverse propagator.
 Finally, the third part specifies the arrangement of time cuts according to the following rule: each possible cut is assigned to the vertex immediately to its left. For each vertex, the set of cuts that follows it, is listed in brackets.

\refstepcounter{section}             
\phantomsection
\section*{Appendix \Alph{section}: Linear reducibility in dynamic diagrams}
\label{AppendA}

In this Appendix, we analyze how the specific structure of dynamic diagrams influences the linear reducibility. 

Obviously, the Feynman representation \eqref{feym_param_1},\eqref{feym_param_2} is suitable for both, static and dynamic diagrams. For a static one (for example, $\phi^4$ model), in the case of massless propagators, $\mathcal{U}$ and $\mathcal{F}$ are linear in all Feynman parameters. Moreover, it has been proven that performing the first four integrations is guaranteed to produce only linear polynomials \cite{Brown08}. However, this property does not hold in the dynamic case, where the presence of cut propagators introduces nonlinear dependencies on the corresponding Feynman parameters in the integrand. To explore this in more detail, let us use an alternative well-known approach to construct this representation by introducing the so-called \textit{Symanzik polynomials} \cite{N_symanzik61}
\begin{equation}
    \mathcal{U} = \det(M)\,,\,\,\,\,\,\,\,\mathcal{F}=\mathcal{U}\left( M_{is} k_i k_s + 2a_i k_i +c \right)\,,
\end{equation}
defined purely by the topology of a diagram. For our purposes, we consider only the first Symanzik polynomial
\begin{equation}
    \label{U_span_trees}
    \mathcal{U} = \sum_T \prod_{e \notin T} u_e\, ,
\end{equation}
where the sum goes over all spanning trees of a diagram. 

As an illustrative example, let us consider the three-loop diagram $Y_3$ from (\ref{3l_new_reduced}) along with its static analogue
\begin{equation}
    \label{G3}
    \vcenter{\hbox{\input{tikz_pics/Y3_numbers}}}  \qquad \quad \vcenter{\hbox{\input{tikz_pics/static_som_numbers}}} \quad .
\end{equation}

\begin{figure}[H]
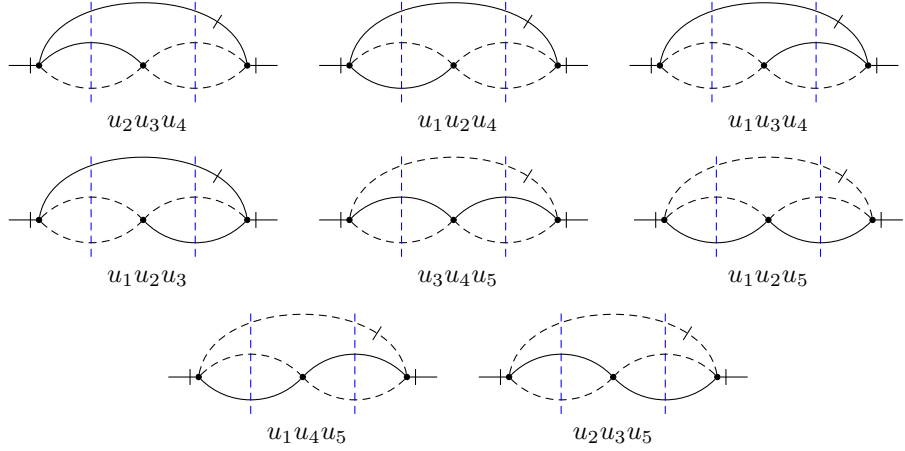

    \centering
    \begin{tabular}{c c c}
    \input{tikz_pics/u2u3u4}     & \input{tikz_pics/u1u2u4} & \input{tikz_pics/u1u3u4} \\
    $u_2 u_3 u_4$     & $u_1 u_2 u_4$ & $u_1 u_3 u_4$ \\
    \input{tikz_pics/u1u2u3} & \input{tikz_pics/u3u4u5} & \input{tikz_pics/u1u2u5}\\
    $u_1 u_2 u_3$ & $u_3 u_4 u_5$ & $u_1 u_2 u_5$
    \end{tabular}
    \begin{tabular}{c c}
    \input{tikz_pics/u1u4u5}     &  \input{tikz_pics/u2u3u5}\\
    $u_1 u_4 u_5$  &  $u_2 u_3 u_5$
    \end{tabular}

    \caption{All spanning trees of diagram $Y_3$ and their corresponding contributions to the first Symanzik polynomial. Black solid lines represent spanning trees; black dotted lines denotes edges not included in the spanning tree; blue dotted lines correspond to dynamic cuts.}
    \label{G3_trees}
\end{figure}

The numbers over the lines and cuts denote the Feynman parameters corresponding to the propagators.  In Fig. \ref{G3_trees}, all contributions to the sum (\ref{U_span_trees}) for the static version are presented. Therefore, its polynomial $\mathcal{U}$ is the following
\begin{equation}
    \mathcal{U}^{\scriptscriptstyle \text{st}}_{\scriptscriptstyle \text{Y3}} = u_2u_3u_4 + u_1u_2u_4 + u_1u_3u_4 + u_1u_2u_3 + u_3u_4u_5 + u_1u_2u_5 + u_1u_4u_5 + u_2u_3u_5\,. 
\end{equation}

Given the expression of $\mathcal{U}^{\scriptscriptstyle \text{st}}_{\scriptscriptstyle \text{Y3}}$ for the static diagram, its dynamic generalization $\mathcal{U}^{\scriptscriptstyle \text{dyn}}_{\scriptscriptstyle \text{Y3}}$ can be obtained by the following substitution of the static parameters $u_i$ by dynamic ones $v_i$ \cite{AVK_RG15}:
\begin{equation}
\label{stat_to_dyn}
    u_i = v_i\,\delta_i + \sum_{cuts}v_j \,, \,\,\,\text{where } \delta_i = \begin{cases}
1\,, & \text{if edge } i \text{ is } \left<\psi \psi\right>, \\
0\,, & \text{if edge } i \text{ is } \left<\psi \psi'\right>.
\end{cases}
\end{equation}
where the sum runs over all indices of the cuts ($6$ or $7$ in our case), which cross the edge $i$. Performing the substitution (\ref{stat_to_dyn}) in these terms yields the first Symanzik polynomial for $Y_3$:

\[
\begin{aligned}
    \mathcal{U}^{\scriptscriptstyle \text{dyn}}_{\scriptscriptstyle \text{Y3}} = & (3 v_2 + 3 v_4 + 6 v_7) v_6^2 + \left(6 v_7^2 + (4 v_1 + 4 v_2 + 4 v_3 + 4 v_4) v_7 + (2 v_2 + 2 v_4) v_1 + (2 v_3 + 2 v_4) v_2 + 2 v_3 v_4\right) v_6 +\\
    &(3 v_1 + 3 v_3) v_7^2 + \left((2 v_2 + 2 v_3 + 2 v_4) v_1 + 2 v_3 (v_2 + v_4)\right) v_7 + \left((v_3 + v_4) v_2 + v_3 v_4\right) v_1 + v_2 v_3 v_4 \\
\end{aligned}
\]

The resulting polynomial is quadratic in the parameters $v_6$ and $v_7$, which correspond to cut propagators. This leads to constraints on the choice of integration order, thus creating challenges for linear reducibility. The nature of these nonlinearities can be readily understood from Fig. \ref{G3_trees}: a spanning tree contributes nonlinearly if any cut crosses more than one edge not included in the tree. The number of such crossed edges determines the degree of nonlinearity. For diagram $Y_3$, every tree contains at least one such nonlinear cut. In general, the higher the loop order of the diagram, the more severe these nonlinearities become. A practical strategy is to eliminate the most problematic cuts during diagram reduction, whenever possible.

\refstepcounter{section}
\phantomsection
\section*{Appendix \Alph{section}: Analytic expressions for diagrams contributions}
\label{AppendB}
We provide the expressions for coefficients $A_{ij}$ in the expansion of the Green function \eqref{gamma_expansion}. $A_{2i}$ and $A_{3i}$ are given analytically, while $A_{41}$ is given only numerically up to 80 digits due to very large analytic contribution from the diagram $C_9$\footnote{The explicit expression for $C_9$ is provided in Supplementary file}. 
\allowdisplaybreaks
\begin{align}
A_{21}\,\,\,\,\, =\,& \frac{2+n}{3}\frac{1}{16}\ln(4/3), \nonumber \\[6pt]
A_{20}\,\,\,\,\, =\,& \frac{2+n}{3}\frac{1}{96} \Big[
8\,G(3,2;1) - 8\,G(0,2;1) + 4\,G(0,3;1) + 24\,\ln^2(2) + 4 (3 + \ln(3)) \ln(2) - 11\,\ln^2(3) - 6\,\ln(3) - 2\,\zeta(2)\Big], \nonumber \\[6pt]
A_{2,-1} =\,& \frac{2+n}{3}\frac{1}{288}\Big[144 \ln(2) - 72 \ln(3) - 48\, G(3, 2, 2; 1) - 72\, G(3, 3, 2; 1) + 48\, G(0, 2, 2; 1) + 24\, G(0, 3, 1; 1) -\nonumber \\
&- 36\, G(0, 3, 3; 1)  + 36\, G(3, 0, 1; 1) - 36\, G(3, 0, 3; 1) + 72\, G(0, 0, 2; 1) + 24\, G(0, 0, 3; 1) + 24\, G(3, 2; 1)- \nonumber \\
& - 24\, G(0, 2; 1) + 12\, G(0, 3; 1) - 33 \ln^{2}(3) + 72 \ln^{2}(2) - \pi^{2} - 35 \ln^{3}(3) + 78 \zeta(3) + 72\, G(3, 2; 1) \ln(2) -\nonumber \\
&- 72\, G(0, 2; 1) \ln(2) + 36\, G(0, 3; 1) \ln(2) + 12 \ln^{2}(2) \ln(3) + 4 \ln(2) \pi^{2} \nonumber \\
&+ 6 \ln(2) \ln^{2}(3) + 12 \ln(2) \ln(3) - 11 \ln(3) \pi^{2} + 140 \ln^{3}(2) \Big], \nonumber\\[6pt]
A_{32}\,\,\,\,\,=\,& \frac{(8+n)(2+n)}{27}\frac{1}{16}\ln(4/3), \nonumber\\[6pt]
A_{31}\,\,\,\,\, =\,& \frac{(8+n)(2+n)}{27}\frac{1}{96} \Big[G(0,3;1)+G(-3,-1;1)+G(0,-3;1)+8G(3,2;1)-8G(0,2;1)-\frac{11}{2}\ln(2)^2+\nonumber\\ &+45\ln(2)\,\ln(3)+24\ln(2) -25\ln(3)^2-12\ln(3)-\frac{5}{12}\pi^2 \Big] , \nonumber\\[6pt]
A_{30}\,\,\,\,\, =\,& \frac{(8+n)(2+n)}{27}\Big[-\frac{27}{64}\ln^{3}(2)+\ln(2)-\frac{9}{32}G(2,-1;1)\ln(2)+\frac{3}{16}G(2,-2;1)\ln(2)+\frac{7}{288}\pi^{2}+\frac{9}{16}G(0,-2,1;1)+\nonumber\\ &+\frac{17}{12}\ln(2)\ln^{2}(3)-\frac{7}{16}G(-1,0,-2;1)+\frac{5}{8}G(0,4;1)\ln(3)-\frac{3}{8}G(2,4;1)\ln(3)+\frac{167}{1152}\ln(2)\pi^{2}+\nonumber\\ &+\frac{25}{48}G(-2,-2,-1;1)+\frac{5}{16}G(-1,-2,-2;1)+\frac{9}{16}G(-2,0,1;1)+\frac{1}{16}G(0,0,2;1)+\frac{1}{16}G(0,-1,2;1)+\nonumber\\ &+\frac{1}{16}G(-1,0,2;1)-\frac{5}{48}G(0,-2,-1;1)-\frac{9}{32}G(0,2;1)\ln(3)-\frac{13}{24}\ln^{3}(3)-\frac{7}{12}G(-1,-2;1)\ln(2)-\nonumber\\ &-\frac{3}{32}G(2,-2;1)\ln(3)+\frac{3}{4}G(2,4;1)\ln(2)+\frac{17}{16}G(0,0,-2;1)+\frac{47}{192}\zeta(3)-\frac{13}{24}G(-2,0,-1;1)-\nonumber\\ &-\frac{5}{8}\ln^{2}(2)\ln(3)+\frac{9}{32}G(-1,1,-2;1)+\frac{11}{16}G(-2,-1;1)\ln(2)-\frac{1}{2}\ln^{2}(3)-\frac{1}{2}\ln(3)+\frac{1}{8}\ln(2)\,G(-2,2;1)-\nonumber\\ &-\frac{17}{16}G(0,-2;1)\ln(3)-\frac{1}{8}G(-1,-2;1)\ln(3)+\frac{17}{32}G(0,2;1)\ln(2)-\frac{1}{8}G(0,-2;1)-\frac{1}{8}G(-1,-2;1)-\nonumber\\ &-\frac{1}{12}G(-2,-1;1)+\frac{1}{2}G(-2,0,-2;1)+\frac{1}{32}G(-1,-1,-2;1)-\frac{1}{16}G(0,-1,-2;1)+\frac{3}{16}G(2,-1;1)\ln(3)-\nonumber\\ &-\frac{9}{16}G(0,1,-2;1)-\frac{119}{1152}\pi^{2}\ln(3)-\frac{5}{48}G(-1,-2,-1;1)+\frac{3}{16}G(-2,-1,-1;1)+\frac{5}{16}G(0,-2,-2;1)-\nonumber\\ &-\frac{11}{24}\ln^{2}(2)+\frac{3}{16}G(-1,2,-1;1)-\frac{19}{16}G(0,4;1)\ln(2)+\frac{53}{48}G(0,-2;1)\ln(2)-\frac{83}{96}G(-2,-1;1)\ln(3)+\nonumber\\ &+\frac{5}{16}G(-2,-1,-2;1)+\frac{23}{24}\ln(2)\ln(3)-\frac{3}{32}G(0,2,-1;1)+\frac{3}{8}G(-2,1,-1;1)\Big] , \nonumber \\[6pt]
A_{41}\,\,\,\,\, =\,\,& 0.22947254878868425251673355790363161890883447986435158174621562622029853848869\,\,\,\,\,\,\,\,+\nonumber \\&0.17118287745692801303776972945966721119333850874317061956175513071212098490820\,\,n\,\:+\nonumber \\&0.03138485130718158671997211268349199393962745933925266631071657688168623479353\,\,n^2+\nonumber \\&0.00158077488794432166513531871478314653508341246687762598319645904035018848080\,\,n^3 . 
\end{align}

\refstepcounter{section}
\phantomsection
\section*{Appendix \Alph{section}: Comparison with the previous analytic three-loop result}
\label{AppendC}
Here we demonstrate the equivalence of the three-loop contribution $b_{31}$ obtained to the renormalization constant $Z_1$
\begin{align}
\label{b_31}
b_{31} =  \frac{1}{96} & \Big[-5G(0,3;1)+G(-3,-1;1)+G(0,-3;1)-4G(3,2;1)+4G(0,2;1)-6\ln(2)-3\ln(3)-\nonumber \\ &-\frac{83}{2}\ln(2)^2+39\ln(2)\ln(3)-\dfrac{17\ln(3)^2}{2}+\dfrac{\pi^2}{12}  \Big]
\end{align}
to the one evaluated in \cite{AV_3l84} for the one-component case $n=1$, which reads
\begin{equation}
\label{b_31_AV}
    b^{(AV)}_{31} = \frac{1}{96}\Big[ 4 \mathrm{Li}_2(3/4) + \ln(4/3) - 26\ln^2(2) -\frac{21}{2}\ln^2(3) + 34 \ln(2)\ln(3) - \frac{1}{2}\pi^2   \Big]\,. \,\,\,\,\,\,\,\,\,\,\,\,\,\,\,\,\,\,\,\,\,\,\,\,\,\,\,\,\,\,\,\,\,\,\,\,
\end{equation}
The result \eqref{b_31} can be re-expressed solely in terms of polylogarithm $\mathrm{Li}_2$ as
\begin{align}
\label{b_31_dilogs}
b_{31} =  \frac{1}{96} & \Big[5\mathrm{Li}_2(1/3)+5\mathrm{Li}_2(-1/2)-5\mathrm{Li}_2(1/4)-\mathrm{Li}_2(-1/3)+6\ln(2)-3\ln(3)-\frac{83}{2}\ln^2(2)-\nonumber \\ &-\frac{17}{2}\ln^2(3)+39\ln(2)\ln(3)+\frac{1}{6}\pi^2  \Big]\,.
\end{align}
The well-known identity
\begin{equation}
    \mathrm{Li}_2(1-x)+\mathrm{Li}_2(1-1/x) = -\frac{1}{2}\ln^2(x)\,\,\,\,\,\,\,\,\,\,\,\,\,
\end{equation}
for polylogarithms in \eqref{b_31_dilogs} yields
\begin{equation}
\label{dilog_rule_1}
    \mathrm{Li}_2(1/3)+\mathrm{Li}_2(-1/2) = -\frac{1}{2}\ln^2(3/2)\,,
\end{equation}
\begin{equation}
\label{dilog_rule_2}
    \mathrm{Li}_2(1/4)=-\mathrm{Li}_2(-1/3)  -\frac{1}{2}\ln^2(4/3)\,.
\end{equation}
Substituting \eqref{dilog_rule_1} and \eqref{dilog_rule_2} into \eqref{b_31_dilogs}, we obtain
\begin{equation}
    \label{b31_compact}
    b_{31} = \frac{1}{96} \Big[ 4\mathrm{Li}_2(-1/3)+3\ln(4/3)-\frac{17}{2}\ln^2(4/3) + \frac{1}{6}\pi^2 \Big] \,.
\end{equation}
This expression is a more compact form of \eqref{b_31_AV}, which can be explicitly derived be using the following relation
\begin{equation}
    \mathrm{Li}_2(1-x) + \mathrm{Li}_2(x) = -\ln(x)\ln(1-x)+\frac{1}{6}\pi^2\,,\,\,\,\,\,\,x<1\,.
\end{equation}
The expression \eqref{b31_compact} is easily generalized to an arbitrary number of components $n$ by multiplying it by ${(n+2)(n+8)}/{27}$.

\refstepcounter{section}
\phantomsection
\section*{Appendix \Alph{section}: Comparison with ${1/n}$\,-\,expansion}
\label{AppendD}
We provide a comparison of the analytic result obtained with the $1/n$-expansion, allowing for partial verification.  In \cite{HHM_2l2}, the leading term of the $1/n$-expansion for the quantity $R\equiv -\gamma^*_{\lambda}/\eta$ was evaluated:
\begin{equation}
    R=\frac{4}{4-d}\Bigg[ \frac{d\,\Gamma(d/2-1)}{8 \Gamma(d-2)\int^{1/2}_0 dx (2x-x^2)^{d/2-2}} -1 \Bigg] + \mathcal{O}(1/n)\,.
\end{equation}
Using the expansion for $\eta$ \eqref{eta_exp}, we derive an expansion for $\gamma_1^*=\eta-\gamma^*_{\lambda}=\eta(1+R)$:
\begin{equation}
\label{gamma_of_s_k}
    \gamma_1^{*}=\frac{8}{n}\bigl(s_2\varepsilon^2+s_3\varepsilon^3+s_4\varepsilon^4+\mathcal{O}(\varepsilon^5)\bigr)+\mathcal{O}\bigl(1/n^2\bigr)\,,
\end{equation}
where
\begin{align}
\label{s_k}
s_2&=1-I_1,\qquad 
s_3=1-I_1-I_2+2I_1^2-\frac{\pi^2}{12},\notag\\[4pt]
s_4&=1-I_1-I_2-I_3+2I_1^2-4I_1^3+4I_1I_2-\frac{\pi^2}{12}+I_1\frac{\pi^2}{6}-\zeta(3)\,,\\
I_n&=\frac{(-1)^n}{n!}\int_{0}^{1/2}\!dx\;\ln^{\,n}\!\bigl[x(2-x)\bigr].
\end{align}
To facilitate comparison with our results, we express $s_k$ in terms of GPLs. Note that $s_4$ is determined solely by the contribution of the diagrams of the topology A, since only these have a structure factor proportional to $n^3$ (see Table \ref{tab:diag_res}).
\begin{align}
 I_1=&\:1-\frac{3}{2}\ln(4/3),\qquad I_2=2-3\ln(4/3)+\frac{1}{4}\ln^{2}\ln(4/3)+2\operatorname{Li}_2(4/3)\,,\notag\\[4pt]
I_3=&\:4G(0,0,-3;1)+4G(0,-3,1;1)-2G(-3,0,-3;1)+2G(-3,0,-1;1) -\notag\\&-2G(0,-3,-3;1)-4G(0,-3;1)+4-6\ln(4/3)
+\frac{1}{2}\ln^{2}(4/3)-\frac{\pi^{2}}{3}\ln(4/3)\,.
\end{align}
Substituting these expressions into \eqref{s_k} yields
\begin{align}
    s_{2} =\:& \frac{3}{2}\ln(4/3), \quad s_{3} = \frac{3}{2}\ln(4/3) - \frac{17}{4}\ln^{2}(4/3) + 2\,\mathrm{Li}_{2}(4/3) + \frac{1}{12}\pi^{2}\,, \label{s2s3} \\ 
    s_{4} =\:& 2G(-3,0,-3;1) - 2G(-3,0,1;1) + 2G(0,-3,-3;1) - 4G(0,-3,1;1) - 4G(0,0,-3;1)+ \nonumber \\ &
 +(12\ln(4/3) - 2)G(0,-3;1) + \Bigl(\frac{\pi^{2}}{12} - \frac{3}{2}\Bigr)\ln(4/3) + \frac{\pi^{2}}{12} - \zeta(3) + \frac{49}{4}\ln^{3}(4/3) - \frac{17}{4}\ln^{2}(4/3)\,. \label{s4}
\end{align}
The expressions \eqref{s2s3} coincide with the result obtained from \eqref{MS},\eqref{our_b31},\eqref{gamma1_simple}, and \eqref{u_star}. However, our expression for $s_4$ appears substantially more elaborate
\begin{align}
\label{s4_our}
s_4 =6\Big[{}& -9\,G(2,0,2;1) + 18\,G(0,1,3;1) - 6\,G(3,2,3;1) - 7\,G(2,-1,1;1) + 12\,G(3,0,2;1) - 21\,G(3,0,1;1) \nonumber\\
& - 48\,G(0,0,2;1) - 52\,G(0,0,3;1) + 6\,G(-1,2,1;1) - 17\,G(0,2,1;1) + 7\,G(0,-1,2;1) + 7\,G(-1,0,2;1) \nonumber\\
& + 10\,G(0,1,2;1) + 12\,G(3,2,2;1) + 12\,G(2,-1,2;1) - G(3,2,1;1) + 13\,G(3,3,2;1) - 9\,G(0,2,2;1) \nonumber\\
& - 34\,G(0,3,1;1) + 12\,G(-1,2,-1;1) - 5\,G(2,0,1;1) - 6\,G(0,2,-1;1) - 12\,G(2,1,3;1) + 3\,G(2,2,3;1) \nonumber\\
& - 16\,G(3,1,2;1) + 9\,G(2,0,3;1) + 6\,G(-1,1,2;1) + 18\,G(0,-2,1;1) + 18\,G(-2,0,1;1) \nonumber\\
& + 12\,G(-2,1,-1;1) + 9\,G(-1,1,-2;1) - 18\,G(0,1,-2;1) - 14\,G(-1,0,-2;1) + 10\,G(-1,-2,-2;1) \nonumber\\
& + 6\,G(-2,-1,-1;1) + 10\,G(-2,-1,-2;1) + G(-1,-1,-2;1) + 34\,G(0,0,-2;1) + 10\,G(0,-2,-2;1) \nonumber\\
& + 16\,G(-2,0,-2;1) - 2\,G(0,-1,-2;1) - \frac{10}{3}\,G(-1,-2,-1;1) - \frac{10}{3}\,G(0,-2,-1;1) \nonumber\\
& - \frac{52}{3}\,G(-2,0,-1;1) + \frac{50}{3}\,G(-2,-2,-1;1) + 3\,G(0,3,3;1) + 3\,G(3,0,3;1) \nonumber\\
& - 2\ln 2\, G(-3,-1;1) + \Bigl(22\ln 2 - \frac{83}{3}\ln 3 - \frac{8}{3}\Bigr) G(-2,-1;1) \nonumber\\
& + \Bigl(-\frac{56}{3}\ln 2 - 4\ln 3 - 4\Bigr) G(-1,-2;1) + \bigl(13\ln 2 - 12\ln 3 - 5\bigr) G(3,2;1) \nonumber\\
& + \Bigl(\frac{106}{3}\ln 2 - 34\ln 3 - 4\Bigr) G(0,-2;1) + \Bigl(5\ln 2 + \frac{3}{2}\ln^2 3 + 5\Bigr) G(0,2;1) \nonumber\\
& + \Bigl(33\ln 2 - \frac{39}{2}\ln 3 + 3\Bigr) G(0,3;1) + \bigl(18\ln 2 - 9\ln 3\bigr) G(-1,-3;1) \nonumber\\
& + \bigl(-16\ln 2 + 12\ln 3\bigr) G(-1,2;1) + \bigl(6\ln 2 - 3\ln 3\bigr) G(2,-2;1) \nonumber\\
& + \bigl(-18\ln 2 + 12\ln 3\bigr) G(2,-1;1) + \Bigl(-6\ln 2 + \frac{9}{2}\ln 3\Bigr) G(2,3;) \nonumber\\
& + \bigl(24\ln 2 - 12\ln 3\bigr) G(2,4;1) + \bigl(-30\ln 2 + 15\ln 3\bigr) G(0,-3;1) \nonumber\\
& + \bigl(-38\ln 2 + 20\ln 3\bigr) G(0,4;1) + 4\ln 2\, G(-2,2;1) \nonumber\\
& - \Bigl(\frac{33}{4}\ln 3 + \frac{20}{3}\Bigr) \ln^2 2
  + \Bigl(\frac{311}{36}\pi^2 + \frac{217}{12}\ln^2 3 + \frac{13}{6}\ln 3 + \frac{1}{2}\Bigr)\ln 2 \nonumber\\
& - \frac{1}{72}\bigl(559\pi^2 + 18\bigr)\ln 3 - \frac{223}{6}\ln^3 2 - \frac{5}{8}\ln^2 3
  - \frac{23}{12}\zeta(3) - \frac{19}{8}\ln^3 3 + \frac{71}{72}\pi^2 \Big] \, .
\end{align}
A numerical evaluation of \eqref{s4} and \eqref{s4_our} confirms that the two expressions are, in fact, identical. The large size difference between these expressions suggests that our results can be greatly simplified.

\vspace{0.2cm}

The result of $1/n$-expansion for $s_4$ \eqref{s4} can be further simplified, analogously to the treatment of $b_{31}$ in Appendix~\ref{AppendC}, using the following relations
\begin{align}
    G(-3,0,-3;1) &= G(-3,0,1;1)+G(0,-3,1;1)+G(0,0,-3;1)\,, \\
    G(0,-3,-3;1) &= G(0,0,-3;1)-G(-3,0,1;1)-\frac{1}{6}\pi^2 \ln(4/3)\,, \\
    G(-3,0,1;1) + &G(0,-3,1;1) = -G(-3,1,0;1)\,,
\end{align}
which yields the following more compact expression
\begin{equation}
\label{s4_1/n}
s_4=\:2G(-3,1,0;1)+\bigl(12\ln(4/3)-2\bigr)\,G(0,-3;1)-\frac{\pi^{2}}{4}\ln(4/3)
+\frac{\pi^{2}}{12}-\zeta(3)-\frac{3}{2}\ln(4/3)-\frac{17}{4}\ln^{2}(4/3)+\frac{49}{4}\ln^{3}(4/3)\,.
\end{equation}

\bibliographystyle{JHEPsortdoi}

\bibliography{main}

\end{document}

%% file: tikz_pics/line5.tex
\begin{tikzpicture}[baseline=(current bounding box.center), node distance=1.5cm]
        \coordinate (v1); 
        \coordinate[right=of v1] (v2);
        \arclls{v1}{v2}{0}{\small $p$}{below};
        \arclls{v1}{v2}{0}{\small \textcolor{blue}{$\varepsilon$}}{above};
    \end{tikzpicture}

%% file: main.bib
@Article{ForsterNelson_Turb_1977,
  author  = {Forster, D. and Nelson, D. R. and Stephen, M. J.},
  title   = {Large‐distance and long‐time properties of a randomly stirred fluid},
  journal = {Phys. Rev. A},
  volume  = {16},
  pages   = {732--745},
  year    = {1977},
  doi     = {10.1103/PhysRevA.16.732}

}

@Book{AVTurbBook1999,
  author    = {Adzhemyan, L.\ Ts. and Antonov, N.\ V. and Vasiliev, A.\ N.},
  title     = {The Field Theoretic Renormalization Group in Fully Developed Turbulence},
  publisher = {Gordon \& Breach},
  year      = {1999},
  isbn      = {9056991450},
  url={https://www.cambridge.org/core/journals/journal-of-fluid-mechanics/article/abs/field-theoretic-renormalization-group-in-fully-developed-turbulence-by-l-ts-adzhemyan-n-v-antonov-a-n-vasiliev-gordon-breach-1999-208-pp-isbn-9056-99145-0-62/037CAAB7814D1CA2EB51B490B4C260C3}
}

@article{AAKVTurb2003,
  title = {Improved $\ensuremath{\epsilon}$ expansion for three-dimensional turbulence: Two-loop renormalization near two dimensions},
  author = {Adzhemyan, L. Ts. and Honkonen, J. and Kompaniets, M. V. and Vasil'ev, A. N.},
  journal = {Phys. Rev. E},
  volume = {71},
  issue = {3},
  pages = {036305},
  numpages = {19},
  year = {2005},
  month = {Mar},
  publisher = {American Physical Society},
  doi = {10.1103/PhysRevE.71.036305},
  url = {https://link.aps.org/doi/10.1103/PhysRevE.71.036305}
}

@Article{CanetTurb22,
  author   = {Canet, L{\'e}onie},
  title    = {Functional renormalisation group for turbulence},
  journal  = {J. Fluid Mech. (Perspectives)},
  volume   = {950},
  pages    = {A1--A20},
  year     = {2022},
  doi      = {10.1017/jfm.2022.808},
  eprint   = {2205.01427},
  archivePrefix = {arXiv},
  primaryClass  = {physics.flu-dyn}
}

@Article{AEKTurb2024,
  author       = {Adzhemyan, L.\ Ts. and Evdokimov, D.\ A. and Kompaniets, M.\ V.},
  title        = {Hyperlogarithms in the theory of turbulence of infinite dimension},
  journal      = {Nucl. Phys. B},
  volume       = {1008},
  pages        = {116716},
  year         = {2024},
  doi          = {10.1016/j.nuclphysb.2024.116716},
  eprint       = {2405.14533},
  archivePrefix = {arXiv},
  primaryClass  = {cond-mat.stat-mech}
}

@article{MZDGG_bact_20,
  author       = {Mahdisoltani, S. and Ben Alì Zinati, R. and Duclut, C. and Gambassi, A. and Golestanian, R.},
  title        = {Nonequilibrium polarity-induced chemotaxis: Emergent Galilean symmetry and exact scaling exponents},
  journal      = {Phys. Rev. Res.},
  volume       = {2},
  pages        = {013100},
  year         = {2020},
  doi          = {10.1103/PhysRevResearch.2.013100},
  eprint       = {1911.08115},
  archivePrefix = {arXiv},
  primaryClass  = {cond-mat.stat-mech}
}

@article{ZDMGG_bact22,
  author       = {Ben Alì Zinati, R. and Duclut, C. and Mahdisoltani, S. and Gambassi, A. and Golestanian, R.},
  title        = {Stochastic dynanics of chemotactic colonies with logistic growth},
  journal      = {EPL},
  volume       = {136},
  pages        = {50003},
  year         = {2022},
  doi          = {10.1209/0295-5075/ac48c9},
  eprint       = {2111.08508},
  archivePrefix = {arXiv},
  primaryClass  = {cond-mat.stat-mech}
}

@article{TT_swarm98,
  author  = {Toner, John and Tu, Yuhai},
  title   = {Flocks, herds, and schools: A quantitative theory of flocking},
  journal = {Phys. Rev. E},
  volume  = {58},
  pages   = {4828--4858},
  year    = {1998},
  doi     = {10.1103/PhysRevE.58.4828}
}

@article{CCGGGP_swarm19,
  author       = {Cavagna, A. and Di Carlo, L. and Giardina, I. and Grandinetti, L. and Grigera, T.\ S. and Pisegna, G.},
  title        = {Dynamical Renormalization Group Approach to the Collective Behavior of Swarms},
  journal      = {Phys. Rev. Lett.},
  volume       = {123},
  pages        = {268001},
  year         = {2019},
  doi          = {10.1103/PhysRevLett.123.268001},
  eprint       = {1905.01227},
  archivePrefix = {arXiv},
  primaryClass = {cond-mat.stat-mech}
}

@article{CCGGMOOS_swarm23,
  author       = {Cavagna, A. and Di Carlo, L. and Giardina, I. and Grigera, T.\ S. and Melillo, S. and Parisi, L. and Pisegna, G. and Scandolo, M.},
  title        = {Natural Swarms in 3.99 Dimensions},
  journal      = {Nat. Phys.},
  volume       = {19},
  pages        = {1043--1049},
  year         = {2023},
  doi          = {10.1038/s41567-023-02028-0}
}

@article{T_swarm24,
  author       = {Toner, John},
  title        = {Birth, Death, and Horizontal Flight: Malthusian flocks with an easy plane in three dimensions},
  journal      = {Phys. Rev. Lett.},
  volume       = {133},
  pages        = {128301},
  year         = {2024},
  doi          = {10.1103/PhysRevLett.133.128301},
  eprint       = {2407.03071},
  archivePrefix = {arXiv},
  primaryClass  = {cond-mat.soft}
}

@article{ZHPL_neural25,
  title={Dynamic neuron approach to deep neural networks: Decoupling neurons for renormalization group analysis},
  author={Lee, Donghee and Lee, Hye-Sung and Yi, Jaeok},
  journal={Phys. Rev. Res.},
  volume={7},
  number={2},
  pages={023276},
  year={2025},
  publisher={APS},
  doi={https://doi.org/10.1103/tf5z-7q29}
}

@article{BC_neural07,
author = {Buice, Michael A. and Cowan, Jack D.},
title = {Field-theoretic approach to fluctuation effects in neural networks},
journal = {Phys. Rev. E},
volume = {75},
pages = {051919},
year = {2007},
doi = {10.1103/PhysRevE.75.051919}
}

@article{LW_neural018,
author = {Li, Shuo-Hui and Wang, Lei},
title = {Neural Network Renormalization Group},
journal = {Phys. Rev. Lett.},
volume = {121},
pages = {260601},
year = {2018},
doi = {10.1103/PhysRevLett.121.260601},
eprint = {1802.02840},
archivePrefix = {arXiv},
primaryClass = {cond-mat.stat-mech}
}

@article{TS_neural22,
author = {Tiberi, Lorenzo and Stapmanns, Jonas and K{"u}hn, Tobias and Luu, Thomas and Dahmen, David and Helias, Moritz},
title = {Gell-Mann--Low Criticality in Neural Networks},
journal = {Phys. Rev. Lett.},
volume = {128},
pages = {168301},
year = {2022},
doi = {10.1103/PhysRevLett.128.168301},
eprint = {2110.01859},
archivePrefix = {arXiv},
primaryClass = {q-bio.NC}
}

@article{GGPS_network25, author = {Gabrielli, Andrea and Garlaschelli, Diego and Patil, Subodh P. and M. {'A}ngeles Serrano}, title = {Network renormalization}, journal = {Nat. Rev. Phys.}, volume = {7}, pages = {203--219}, year = {2025}, doi = {10.1038/s42254-025-00817-5}, eprint = {2412.12988}, archivePrefix = {arXiv}, primaryClass = {cond-mat.stat-mech} }

@article{NW_network99, author = {Newman, M. E. J. and Watts, D. J.}, title = {Renormalization group analysis of the small-world network model}, journal = {Phys. Lett. A}, volume = {263}, pages = {341--346}, year = {1999}, doi = {10.1016/S0375-9601(99)00757-4}, eprint = {cond-mat/9903357}, archivePrefix = {arXiv}, primaryClass = {cond-mat.stat-mech} }

@article{KJBS_network24,
author = {van der Kolk, Jasper and Bogu{~n}'a, Mari{'a}n and Serrano, M. {'A}ngeles},
title = {Renormalization of networks with weak geometric coupling},
journal = {Phys. Rev. E},
volume = {110},
pages = {L032302},
year = {2024},
doi = {10.1103/PhysRevE.110.L032302},
eprint = {2403.12663},
archivePrefix = {arXiv},
primaryClass = {physics.soc-ph}
}

@article{DL_vapour84,
author = {van Dieren, F. and van Leeuwen, J. M. J.},
title = {Renormalization of the gas-liquid transition: I. The flow equations},
journal = {Physica A},
volume = {128},
pages = {383--403},
year = {1984},
doi = {10.1016/0378-4371(84)90182-1}
}

@article{DL_vapour86,
author = {van Dieren, F. and van Leeuwen, J. M. J.},
title = {Renormalization of the gas-liquid transition: II. Results for some renormalization schemes},
journal = {Physica A},
volume = {136},
pages = {21--46},
year = {1986},
doi = {10.1016/0378-4371(86)90041-5}
}

@article{SHH_vapour76,
author = {Siggia, E. D. and Halperin, B. I. and Hohenberg, P. C.},
title = {Renormalization-group treatment of the critical dynamics of the binary-fluid and gas-liquid transitions},
journal = {Phys. Rev. B},
volume = {13},
pages = {2110--2136},
year = {1976},
doi = {10.1103/PhysRevB.13.2110}
}

@article{AVKK_Hmodel99,
author = {Adzhemyan, L. Ts. and Vasiliev, A. N. and Kabrits, Yu. S. and Kompaniets, M. V.}, 
title = {H-model of critical dynamics: Two-loop calculations of RG functions and critical indices}, 
journal = {Theor. Math. Phys.}, volume = {119}, number = {1}, pages = {454--470}, year = {1999}, doi = {10.1007/BF02557344} }

@book{Vasiliev_book,
author = {Vasil'ev, A. N.},
title = {The Field Theoretic Renormalization Group in Critical Behavior Theory and Stochastic Dynamics},
publisher = {Chapman \& Hall/CRC},
address = {London},
year = {2004},
isbn = {978-0-415-31002-4},
doi={https://doi.org/10.1201/9780203483565-9}
}

@book{ZinnJusti_book,
author = {Zinn-Justin, Jean},
title = {Quantum Field Theory and Critical Phenomena},
publisher = {Clarendon Press, Oxford University Press},
address = {Oxford},
year = {2002},
edition = {4th},
doi = {10.1093/acprof:oso/9780198509233.001.0001},
isbn = {0198509235}
}

@article{Fisher1998, author = {Fisher, M. E.}, title = {Renormalization group theory: Its basis and formulation in statistical physics}, journal = {Rev. Mod. Phys.}, volume = {70}, pages = {653--681}, year = {1998}, doi = {10.1103/RevModPhys.70.653} }

@article{Halperin77,
  title={Theory of dynamic critical phenomena},
  author={Hohenberg, Pierre C and Halperin, Bertrand I},
  journal={Rev. Mod. Phys.},
  volume={49},
  number={3},
  pages={435},
  year={1977},
  publisher={APS},
  doi={https://doi.org/10.1103/RevModPhys.49.435}
}

@article{LV_ising23,
  title={Critical dynamical behavior of the Ising model},
  author={Liu, Zihua and Vatansever, Erol and Barkema, Gerard T and Fytas, Nikolaos G},
  journal={Phys. Rev. E},
  volume={108},
  number={3},
  pages={034118},
  year={2023},
  publisher={APS},
  doi={https://doi.org/10.1103/physreve.108.034118}
}

@article{AEHIKKZ_5l22,
  title={The dynamic critical exponent z for 2d and 3d Ising models from five-loop $\varepsilon$ expansion},
  author={Adzhemyan, L Ts and Evdokimov, DA and Hnati{\v{c}}, M and Ivanova, EV and Kompaniets, MV and Kudlis, A and Zakharov, DV},
  journal={Phys. Lett. A},
  volume={425},
  pages={127870},
  year={2022},
  publisher={Elsevier},
  doi={https://doi.org/10.1016/j.physleta.2021.127870}
}

@article{BHS_XY01,
  title={Nonequilibrium critical dynamics of the two-dimensional XY model},
  author={Berthier, Ludovic and Holdsworth, Peter CW and Sellitto, Mauro},
  journal={J. Phys. A: Math. Theor.},
  volume={34},
  number={9},
  pages={1805},
  year={2001},
  publisher={IOP Publishing},
  doi={https://doi.org/10.1088/0305-4470/34/9/301}
}

@article{J_XY00,
  title={Critical dynamics of the four-dimensional XY model},
  author={Jensen, Lars Melwyn and Kim, Beom Jun and Minnhagen, Petter},
  journal={Physica B},
  volume={284},
  pages={455--456},
  year={2000},
  publisher={Elsevier},
  doi={https://doi.org/10.1016/s0921-4526(99)02024-4}
}

@article{AR_heisenberg25,
  title={Equilibrium and out-of-equilibrium critical dynamics of the three-dimensional Heisenberg model with random cubic anisotropy},
  author={Astillero, A and Ruiz-Lorenzo, JJ},
  journal={Phys. Rev. E},
  volume={111},
  number={6},
  pages={064126},
  year={2025},
  publisher={APS},
  doi={https://doi.org/10.1103/q72t-1xw7}
}

@article{PS_Heisenberg89,
  title={Critical dynamics in Heisenberg ferromagnets and antiferromagnets near the percolation threshold},
  author={Pimentel, IR and Stinchcombe, RB},
  journal={J. Phys. A: Math. Theor.},
  volume={22},
  number={18},
  pages={3959},
  year={1989},
  publisher={IOP Publishing},
  doi={https://doi.org/10.1088/0305-4470/22/18/029}
}

@article{Nalimov_superfluid19,
  title={Critical dynamics of the phase transition to the superfluid state},
  author={Zhavoronkov, Yu A and Komarova, Marina Vladimirovna and Molotkov, Yu G and Nalimov, M Yu and Honkonent, J},
  journal={Theor. Math. Phys.},
  volume={200},
  number={2},
  pages={1237--1251},
  year={2019},
  publisher={Springer},
  doi={https://doi.org/10.1134/s0040577919080142}
}

@article{Nalimov_superfluid23,
  title={Composite operators of stochastic model A},
  author={Davletbaeva, Diana and Hnati{\v{c}}, M and Komarova, Marina Vladimirovna and Lu{\v{c}}ivjansk{\`y}, Tom{\'a}{\v{s}} and Mi{\v{z}}i{\v{s}}in, Lukas and Nalimov, M Yu},
  journal={Theor. Math. Phys.},
  volume={216},
  number={3},
  pages={1349--1359},
  year={2023},
  publisher={Springer},
  doi={https://doi.org/10.1134/s004057792309009x}
}

@article{NG_multiferroic15,
  title={Critical slowing down near the multiferroic phase transition in $MnWO_4$},
  author={Niermann, D and Grams, CP and Becker, P and Bohat{\`y}, L and Schenck, H and Hemberger, J},
  journal={Phys. Rev. Lett.},
  volume={114},
  number={3},
  pages={037204},
  year={2015},
  publisher={APS},
  doi={https://doi.org/10.1103/physrevlett.114.037204}
}

@article{L_alloy15,
  title={Ordering fluctuation dynamics in $AuAgZn_2$},
  author={Livet, Fr{\'e}d{\'e}ric and F{\`e}vre, Mathieu and Beutier, Guillaume and Sutton, Mark},
  journal={Phys. Rev. B},
  volume={92},
  number={9},
  pages={094102},
  year={2015},
  publisher={APS},
  doi={https://doi.org/10.1103/physrevb.92.094102}
}

@article{AT_ion25,
  title={Renormalized critical dynamics and fluctuations in model A in the Hohenberg-Halperin classification},
  author={Attieh, Nadine and Touroux, Nathan and Bluhm, Marcus and Kitazawa, Masakiyo and Sami, Taklit and Nahrgang, Marlene},
  journal={Phys. Rev. C},
  volume={111},
  number={2},
  pages={024906},
  year={2025},
  publisher={APS},
  doi={https://doi.org/10.1103/physrevc.111.024906}
}

@article{AEHIKKZ_5l_2_22,
  title={Model A of critical dynamics: 5-loop $\varepsilon$ expansion study},
  author={Adzhemyan, L Ts and Evdokimov, DA and Hnati{\v{c}}, M and Ivanova, EV and Kompaniets, MV and Kudlis, A and Zakharov, DV},
  journal={Physica A},
  volume={600},
  pages={127530},
  year={2022},
  publisher={Elsevier},
  doi={10.1016/j.physa.2022.127530}
}

@article{O_lattice04,
  title={Universality classes in nonequilibrium lattice systems},
  author={{\'O}dor, G{\'e}za},
  journal={Rev. Mod. Phys.},
  volume={76},
  number={3},
  pages={663--724},
  year={2004},
  publisher={APS},
  doi={https://doi.org/10.1103/RevModPhys.76.663}
}

@article{AV_3l84,
  title={Critical dynamics as a field theory},
  author={Antonov, N V and Vasil'ev, A N},
  journal={Theor. Math. Phys.},
  volume={60},
  pages={671–679},
  year={1984},
  doi={https://doi.org/10.1007/BF01018251}
}

@article{AIKV_4lSD17,
  title={Diagram reduction in problem of critical dynamics of ferromagnets: 4-loop approximation},
  author={Adzhemyan, L Ts and Ivanova, EV and Kompaniets, MV and Vorobyeva, S Ye},
  journal={J. Phys. A: Math. Theor.},
  volume={51},
  number={15},
  pages={155003},
  year={2018},
  publisher={IOP Publishing},
  doi={https://doi.org/10.1088/1751-8121/aab20f}
}

@article{HHM_2l2,
  title={Calculation of dynamic critical properties using Wilson's expansion methods},
  author={Halperin, BI and Hohenberg, PC and Ma, Shang-keng},
  journal={Phys. Rev. Lett.},
  volume={29},
  number={23},
  pages={1548},
  year={1972},
  publisher={APS},
  doi={https://doi.org/10.1103/physrevlett.29.1548}
}

@article{ANS_4l08,
  title={Calculation of dynamical exponent in model A of critical dynamics to order $\varepsilon^4$},
  author={Adzhemyan, L. Ts. and  Novikov, S. V. and Sladkoff, L. },
  journal={Vestn.St.Petersbg.Univ., Phys.Chem},
  number={4},
  pages={110-114},
  year={2008},
  doi={https://doi.org/10.48550/arXiv.0808.1347}
}

@article{HKN_hoa05,
  title={Large-order asymptotes for dynamic models near equilibrium},
  author={Honkonen, Juha and Komarova, MV and Nalimov, M Yu},
  journal={Nucl. Phys. B},
  volume={707},
  number={3},
  pages={493--508},
  year={2005},
  publisher={Elsevier},
  doi={https://doi.org/10.1016/j.nuclphysb.2004.11.016}
}

@article{Schnetz_7l22,
  title={$\phi^4$ theory at seven loops},
  author={Schnetz, Oliver},
  journal={Phys. Rev. D},
  volume={107},
  number={3},
  pages={036002},
  year={2023},
  publisher={APS},
  doi={10.1103/PhysRevD.107.036002}
}

@inproceedings{ADHIK_multi16,
  title={Multi-loop calculations of anomalous exponents in the models of critical dynamics},
  author={Adzhemyan, L Ts and Dan{\v{c}}o, M and Hnati{\v{c}}, M and Ivanova, EV and Kompaniets, MV},
  booktitle={EPJ Web of Conferences},
  volume={108},
  pages={02004},
  year={2016},
  organization={EDP Sciences},
  doi={https://doi.org/10.1051/epjconf/201610802004}
}

@article{IKKN_multi23,
  title={Quantum-field multiloop calculations in critical dynamics},
  author={Ivanova, Ella and Kalagov, Georgii and Komarova, Marina and Nalimov, Mikhail},
  journal={Symmetry},
  volume={15},
  number={5},
  pages={1026},
  year={2023},
  publisher={MDPI},
  doi={https://doi.org/10.3390/sym15051026}
}

@article{Brown08,
  title={The massless higher-loop two-point function},
  author={Brown, Francis},
  journal={Commun. Math. Phys.},
  volume={287},
  number={3},
  pages={925--958},
  year={2009},
  publisher={Springer},
  doi={https://doi.org/10.1007/s00220-009-0740-5},
  
}

@article{Panzer15,
  title={Algorithms for the symbolic integration of hyperlogarithms with applications to Feynman integrals},
  author={Panzer, Erik},
  journal={Comput. Phys. Commun.},
  volume={188},
  pages={148--166},
  year={2015},
  publisher={Elsevier},
  doi={https://doi.org/10.1016/j.cpc.2014.10.019}
}

@article{BGKS_phi321,
  title={Five-loop renormalization of $\phi^3$ theory with applications to the Lee-Yang edge singularity and percolation theory},
  author={Borinsky, Michael and Gracey, John A and Kompaniets, Mikhali V and Schnetz, Oliver},
  journal={Phys. Rev. D},
  volume={103},
  number={11},
  pages={116024},
  year={2021},
  publisher={APS},
  doi={https://doi.org/10.1103/physrevd.103.116024}
}

@article{KP17,
  title={Minimally subtracted six-loop renormalization of $O(n)$-symmetric $\phi^4$ theory and critical exponents},
  author={Kompaniets, Mikhail V and Panzer, Erik},
  journal={Phys. Rev. D},
  volume={96},
  number={3},
  pages={036016},
  year={2017},
  publisher={APS},
  doi={https://doi.org/10.1103/physrevd.96.036016}
}

@article{KP16,
    author = {Kompaniets, Mikhail V and Panzer, Erik},
    title = {Renormalization group functions of $\phi^4$ theory in the MS-scheme to six loops},
    primaryClass = "hep-th",
    doi = "10.22323/1.260.0038",
    journal = "PoS",
    volume = "LL2016",
    pages = "038",
    year = "2016"
}

@article{RatioRoots,
  title={RationalizeRoots: software package for the rationalization of square roots},
  author={Besier, Marco and Wasser, Pascal and Weinzierl, Stefan},
  journal={Comput. Phys. Commun.},
  volume={253},
  pages={107197},
  year={2020},
  publisher={Elsevier},
  doi={https://doi.org/10.1016/j.cpc.2020.107197}
}

@article{Panzer_lr14,
    author = {Panzer, Erik},
    title = {On hyperlogarithms and Feynman integrals with divergences and many scales},
    doi = {10.1007/JHEP03(2014)071},
    journal = {JHEP},
    volume = {03},
    pages = {071},
    year = {2014}
}

@article{HM_LR20,
  title={Multiple polylogarithms with algebraic arguments and the two-loop EW-QCD Drell-Yan master integrals},
  author={Heller, Matthias and von Manteuffel, Andreas and Schabinger, Robert M},
  journal={Phys. Rev. D},
  volume={102},
  number={1},
  pages={016025},
  year={2020},
  publisher={APS},
  doi={https://doi.org/10.1103/physrevd.102.016025}
}

@article{BPS_hep20,
    author = {Bonetti, Marco and Panzer, Erik and Smirnov, Vladimir A. and Tancredi, Lorenzo},
    title = {Two-loop mixed QCD-EW corrections to $gg \to Hg$},
    eprint = {2007.09813},
    archivePrefix = {arXiv},
    primaryClass = {hep-ph},
    doi = {10.1007/JHEP11(2020)045},
    journal = {JHEP},
    volume = {11},
    pages = {045},
    year = {2020}
}

@article{CS_hep22,
  title={Planar three-loop master integrals for 2 → 2 processes with one external massive particle},
  author={Canko, Dhimiter D and Syrrakos, Nikolaos},
  journal={JHEP},
  volume={2022},
  number={4},
  pages={1--19},
  year={2022},
  publisher={Springer},
  doi={https://doi.org/10.1007/jhep04(2022)134}
}

@article{FL_hep23,
  title={Massive three-loop form factors: Anomaly contribution},
  author={Fael, Matteo and Lange, Fabian and Sch{\"o}nwald, Kay and Steinhauser, Matthias},
  journal={Phys. Rev. D},
  volume={107},
  number={9},
  pages={094017},
  year={2023},
  publisher={APS},
  doi={https://doi.org/10.1103/physrevd.107.094017}
}

@article{LZ_hep22,
  title={Master integrals for mixed QCD-QED corrections to charged-current Drell-Yan production of a massive charged lepton},
  author={Long, Ming-Ming and Zhang, Ren-You and Ma, Wen-Gan and Jiang, Yi and Han, Liang and Li, Zhe and Wang, Shuai-Shuai},
  journal={JHEP},
  volume={2022},
  number={7},
  pages={1--35},
  year={2022},
  publisher={Springer},
  doi={https://doi.org/10.1007/jhep07(2022)078}
}

@article{MP_hep20,
  title={Cusp and collinear anomalous dimensions in four-loop QCD from form factors},
  author={von Manteuffel, Andreas and Panzer, Erik and Schabinger, Robert M},
  journal={Phys. Rev. Lett.},
  volume={124},
  number={16},
  pages={162001},
  year={2020},
  publisher={APS},
  doi={https://doi.org/10.1103/physrevlett.124.162001}
}

@article{AM_hep21,
  title={Four-loop collinear anomalous dimensions in QCD and N= 4 super Yang-Mills},
  author={Agarwal, Bakul and von Manteuffel, Andreas and Panzer, Erik and Schabinger, Robert M},
  journal={Phys. Lett. B},
  volume={820},
  pages={136503},
  year={2021},
  publisher={Elsevier},
  doi={https://doi.org/10.1016/j.physletb.2021.136503}
}

@article{BH_sd04,
  title={Numerical evaluation of multi-loop integrals by sector decomposition},
  author={Binoth, Thomas and Heinrich, G},
  journal={Nucl. Phys. B},
  volume={680},
  number={1-3},
  pages={375--388},
  year={2004},
  publisher={Elsevier},
  doi={https://doi.org/10.1016/j.nuclphysb.2003.12.023}
}

@article{panzerPhD15,
  title={Feynman integrals and hyperlogarithms},
  author={Panzer, Erik},
  journal={PhD thesis},
  eprint = "1401.4361",
  archivePrefix = "arXiv",
  year={2015},
}

@article{Brown13,
  title={Angles, scales and parametric renormalization},
  author={Brown, Francis and Kreimer, Dirk},
  journal={Lett. Math. Phys.},
  volume={103},
  number={9},
  pages={933--1007},
  year={2013},
  publisher={Springer},
  doi={https://doi.org/10.1007/s11005-013-0625-6}
}

@article{Brown09,
  title={On the periods of some Feynman integrals},
  author={Brown, Francis},
  journal={arXiv preprint arXiv:0910.0114},
  year={2009},
  url={https://arxiv.org/abs/0910.0114}
}

@article{V_irr8,
  title={Methods of calculating many-loop diagrams and renormalization-group analysis of the $\phi^4$ theory},
  author={Vladimirov, AA},
  journal={Theor. Math. Phys.},
  volume={36},
  number={2},
  pages={732--737},
  year={1978},
  publisher={Springer},
  doi={https://doi.org/10.1007/bf01036487}
}

@article{chetyrkin1980,
  title={New approach to evaluation of multiloop Feynman integrals: The Gegenbauer polynomial x-space technique},
  author={Chetyrkin, KG and Kataev, AL and Tkachov, FV},
  journal={Nucl. Phys. B},
  volume={174},
  number={2-3},
  pages={345--377},
  year={1980},
  publisher={Elsevier},
  doi={https://doi.org/10.1016/0550-3213(80)90289-8}
}

@article{R_shuffle58,
  title={Lie elements and an algebra associated with shuffles},
  author={Ree, Rimhak},
  journal={Ann. Math.},
  volume={68},
  number={2},
  pages={210--220},
  year={1958},
  publisher={JSTOR},
  doi={https://doi.org/10.2307/1970243}
}

@article{EllipHpl18,
  title={Elliptic polylogarithms and iterated integrals on elliptic curves. Part I: general formalism},
  author={Johannes Broedel and Claude Duhr and Falko Dulat and Lorenzo Tancredi},
  journal={JHEP},
  volume={93},
  year={2018},
  doi={https://doi.org/10.1007/JHEP05(2018)093}
}

@article{EllipHpl19,
  title={Elliptic polylogarithms and Feynman parameter integrals},
  author={Broedel, Johannes and Duhr, Claude and Dulat, Falko and Penante, Brenda and Tancredi, Lorenzo},
  journal={JHEP},
  volume={2019},
  number={5},
  pages={1--38},
  year={2019},
  publisher={Springer},
  doi={https://doi.org/10.1007/jhep05(2019)120}
}

@article{BY_K318,
  title={Traintracks through Calabi-Yau manifolds: scattering amplitudes beyond elliptic polylogarithms},
  author={Bourjaily, Jacob L and He, Yang-Hui and McLeod, Andrew J and Von Hippel, Matt and Wilhelm, Matthias},
  journal={Phys. Rev. Lett.},
  volume={121},
  number={7},
  pages={071603},
  year={2018},
  publisher={APS},
  doi={https://doi.org/10.1103/physrevlett.121.071603}
}

@article{D_K325,
  title={Feynman integrals, elliptic integrals and two-parameter K3 surfaces},
  author={Duhr, Claude and Maggio, Sara},
  journal={JHEP},
  volume={2025},
  number={6},
  pages={1--24},
  year={2025},
  publisher={Springer},
  doi = {https://doi.org/10.1007/JHEP06(2025)250}
}

@article{AVK_RG15,
  title={Representation of the $\beta$-function and anomalous dimensions by nonsingular integrals in models of critical dynamics},
  author={Adzhemyan, L. T. and Vorob'eva, S. E. and Kompaniets, M. V.},
  journal={Theor. Math. Phys.},
  volume={185},
  pages={1361-1369},
  year={2015},
  doi={10.1007/s11232-015-0345-4}
}

@article{N_symanzik61,
  title={Parametric integral formulas and analytic properties in perturbation theory},
  author={Nakanishi, Noboru},
  journal={Prog. Theor. Phys. Suppl.},
  volume={18},
  pages={1--81},
  year={1961},
  publisher={Oxford University Press},
  doi={https://doi.org/10.1143/PTPS.18.1}
}

@article{Ginac1,
  author = "Christian W. Bauer and Alexander Frink and Richard B. Kreckel",
  title = "{Introduction to the GiNaC Framework for Symbolic Computation within the C++ Programming Language}",
  journal = "J. Symb. Comput.",
  volume = "33",
  number = "1",
  pages = "1--12",
  year = "2002",
  doi = "10.1006/jsco.2001.0494",
  archivePrefix = "arXiv",
  eprint = "cs/0004015",
  primaryClass = "cs.SC"
}

@article{Ginac2,
  author = "Jens Vollinga and Stefan Weinzierl",
  title = "{Numerical evaluation of multiple polylogarithms}",
  journal = "Comput. Pyhs. Commun.",
  volume = "167",
  number = "3",
  pages = "177--194",
  year = "2005",
  doi = "10.1016/j.cpc.2004.12.009",
  archivePrefix = "arXiv",
  eprint = "hep-ph/0410259",
  primaryClass = "hep-ph"
}

@article{G_symbol10,
  title={Classical polylogarithms for amplitudes and Wilson loops},
  author={Goncharov, Alexander B and Spradlin, Marcus and Vergu, C and Volovich, Anastasia},
  journal={Phys. Rev. Let.},
  volume={105},
  number={15},
  pages={151605},
  year={2010},
  publisher={APS},
  doi={https://doi.org/10.1103/PhysRevLett.105.151605}
}

@article{D12_symbol,
  title={Hopf algebras, coproducts and symbols: an application to Higgs boson amplitudes},
  author={Duhr, Claude},
  journal={JHEP},
  volume={2012},
  number={8},
  pages={1--46},
  year={2012},
  publisher={Springer},
  doi={10.1007/JHEP08(2012)043}
}

@article{D19_symbol,
    author = "Duhr, Claude and Dulat, Falko",
    title = "{PolyLogTools {\textemdash} polylogs for the masses}",
    eprint = "1904.07279",
    archivePrefix = "arXiv",
    primaryClass = "hep-th",
    reportNumber = "CP3-19-17, CERN-TH-2019-045, SLAC-PUB-17423",
    doi = "10.1007/JHEP08(2019)135",
    journal = "JHEP",
    volume = "08",
    pages = "135",
    year = "2019"
}

@article{N_nickel1,
  title={On ordering and identifying undirected linear graphs},
  author={Nagle, John F},
  journal={J. Math. Phys.},
  volume={7},
  number={9},
  pages={1588--1592},
  year={1966},
  publisher={AIP Publishing},
  doi={https://doi.org/10.1063/1.1705069}
}

@article{N_nickel2,
  title={Compilation of 2-pt and 4-pt graphs for continuous spin model},
  author={Nickel, BG and Meiron, DI and Baker Jr, GA},
  journal={University of Guelph report},
  year={1977}
}

@book{ChengWu,
    author = {Cheng, Hung and Wu, T. T.},
    title = {Expanding protons: scattering at high-energies},
    publisher={MIT Press},
    year = {1987}
}

@article{Perc25,
  title = {Field-theoretic analysis of dynamic isotropic percolation: Three-loop approximation},
  author = {Hnati\ifmmode \check{c}\else \v{c}\fi{}, Michal and Kecer, Matej and Kompaniets, Mikhail V. and Lu\ifmmode \check{c}\else \v{c}\fi{}ivjansk\'y, Tomas and Mi\ifmmode \check{z}\else \v{z}\fi{}i\ifmmode \check{s}\else \v{s}\fi{}in, Lukas and Molotkov, Yurii G.},
  journal = {Phys. Rev. E},
  volume = {112},
  issue = {1},
  pages = {014113},
  numpages = {14},
  year = {2025},
  doi = {10.1103/87m5-mt2b}
}

@manual{maple,
  title        = {Maple},
  author       = {{Maplesoft}},
  organization = {Maplesoft, a division of Waterloo Maple Inc.},
  address      = {Waterloo, Ontario},
  note         = {Maple Computer Algebra System},
}

@article{GraphState14,
    author = "Batkovich, D. and Kirienko, Yu. and Kompaniets, M. and Novikov, S.",
    title = "{GraphState - a tool for graph identification and labelling}",
    eprint = "1409.8227",
    archivePrefix = "arXiv",
    primaryClass = "hep-ph",
    year = "2014"
}
